\documentclass{article}
\usepackage[utf8]{inputenc}
\usepackage{authblk}
\usepackage{setspace}
\usepackage[margin=1.25in]{geometry}
\usepackage{graphicx}
\usepackage{subcaption}
\usepackage{amsmath}
\usepackage{lineno}
\usepackage{float}
\usepackage{xcolor}

\usepackage[style=nejm, 
citestyle=numeric-comp,
sorting=none]{biblatex}
\title{Test mass charge management in the detection of gravitational waves in space based on UV micro-LED}

\author[1]{Yuandong Jia}
\author[2]{Zhihao Zhang}
\author[1]{Yinbowen Zhang}
\author[2]{Yuning Gu}
\author[3]{Suwen Wang}
\author[1]{Guozhi Chai}
\author[1]{Zemin Zhang}
\author[4]{Yi Zhang}
\author[2]{Shanduan Zhang}
\author[1*]{Hongqing Huo}
\author[5**]{Zongfeng Li}
\author[2***]{Pengfei Tian}
\author[6]{Yun Kau Lau}

\affil[1]{School of Physical Science and Technology, Lanzhou University, Lanzhou, China}
\affil[2]{School of Information Science and Technology, Fudan University, Shanghai, China}
\affil[3]{Ocean Observation Institute of Technology, Hainan Tropical Ocean University, Sanya, China}
\affil[4]{School of Nuclear Science and Technology, Lanzhou University, Lanzhou, China}

\affil[5]{Key Laboratory of Space Utilization, Technology and Engineering Center for Space Utilization, Chinese Academy of Sciences, Beijing,China}
\affil[6]{Institute of Applied Mathematics, Morningside Center of Mathematics, Academy of Mathematics and System Science, Chinese Academy of Sciences, Beijing, China}
\affil[*]{Address correspondence to: huohq@lzu.edu.cn}
\affil[**]{Address correspondence to: lzfeng@csu.ac.cn}
\affil[***]{Address correspondence to: pftian@fudan.edu.cn}

\begin{document}

\maketitle

\begin{abstract}
As an alternative to the ultraviolet light emitting diode(UV LED), the feasibility of utilizing UV micro-LED in the charge management in the detection of gravitational waves in space is experimentally studied. Compared with UV LED, micro-LED is more compact in size, has better current spreading, faster response time and longer operating life. Performance characteristics of micro-LEDs were measured, with peak wavelength of 254 nm, 262 nm, 274 nm, and 282 nm for each respective micro-LED, and the photoelectric effect was demonstrated. The effectiveness of micro-LED based charge management experiments were demonstrated using above micro-LEDs mounted on a cubical test mass, and different discharge rates were achieved by varying the drive current and duty cycle using pulse width modulation(PWM). Laboratory data was also shown to demonstrate the space qualification of the micro-LED device, the key electrical and optical characteristics of the micro-LEDs showed less than 5{\%} variation. The results of the qualification bring the micro-LED device Technology Readiness Level(TRL) to TRL-5. TRL-6 will be reached provided additional radiation and thermal tests are conducted and in a position ready to be flown and further tested in space.
\end{abstract}

\section{Introduction}

In the detection of gravitational waves in space, spacecrafts are exposed to the heliospheric space environment and subject to continuous bombardment by high energy particles coming from the galactic cosmic rays and solar activity. From our simulation study and a number of other studies, positively charged particles like protons and alpha particles with energy higher than 80 MeV will penetrate a spacecraft and deposit in a test mass in the interior spacecraft. If not properly controlled, the electrical noise generated by a charged test mass will overwhelm the gravitational wave signals and jeopardise the entire mission. An open loop charge control system is needed to neutralise the charges and ensures that the inertial sensor functions properly as controller for the dragfree control loop. 

Beginning from the Gravity Probe B(GP-B) mission and later on the Laser Interferometer Space Antenna(LISA) Pathfinder, photoelectric effect is employed as a way to neutralise charges in a test mass\cite{CMC, LLP}. The test mass or the electrode housing enclosing it are irradiated by ultra violent light and electrons are ejected from both surfaces by means of the photoelectric effect. This enables us to keep the electrical noise of a test mass down to a certain level in terms of residual acceleration noise and at the same time no contact with the test mass is made. In the case of GP-B and LISA Pathfinder, a mercury lamp was used to generate the ultraviolent light source. In the upcoming LISA and the prospective Chinese missions, UV LED  light source is considered as a better alternative than a mercury lamp for a number of technical reasons\cite{CTD, TTP, TAS}. Under the auspices of the gravitational wave program of the National Science Council, China, the Huazhong gravity group pioneer the study of micro-LED as an alternative to UV LED as the light source for charge management, with attention mainly confined to the standard 255 nm in peak wavelength\cite{HEDU, MEUD}. 

Micro-LED is widely used in display, optical communication, direct writing, biomedical, and other applications\cite{VSIB, 2GF, AAM, MNP}. As far as detection of gravitational waves in space is concerned, compared with UV LED, micro-LED is more compact in size and weight, with better resolution in optical power down to the pW level. A micro lens system also enables us to steer a light beam in the appropriate direction to optimize the efficiency of photoelectric effect\cite{GMDA}. Further, the prospect of integrating the micro-LED to be part of the structure of the electrode housing for a test mass, without directing light into the housing by means of fibre that results in optical loss, prompts us to look more indepth of the micro-LED as a possible alternative to UV LED. Pengfei Tian's group at Fudan University, China has been engaged in micro-LED research for a long time, lately also conducted relevant research on the coupling between micro-LEDs and optical fibers in the charge management system\cite{10GVL, RIMS, HEDU}. Micro-LEDs used in this paper are manufactured by Tian's group.

The present work is structured as follows. In section 2, measurement data on the characteristics of micro-LED relevant for charge management is given. In section 3, we will demonstrate the photoelectric effect for micro-LEDs with wavelengths different from the standard 255 nm ultraviolent light. The study will enable us to understand better the optimization of photoelectric efficiency for scenarios in different phases of a solar cycle. Continuous charge management using micro-LED will be studied in section 4, in the expectation that a finer optical power resolution of micro-LED will enable us to further minimise the residual charge level when when we have a better modelling of galactic cosmic rays when solar activity is less violent. Further, so far micro-LED is used on ground for a number of purposes not related to space. It is not known whether this light source is suitable for space applications. The key part of this work in section 5 is the space qualification of micro-LED and raise its technological readiness level to TRL-5, close to  a position to be ready to fly in space. Some remarks to look ahead to future work will be made in the final section to conclude the present work.

\section{Performance characteristics of micro-LEDs}

In this section, data on the measurements on the characteristics of micro-LED relevant for charge management will be given. This paves the way for further experimental investigations, to be presented in the next few sections. 

\begin{figure}[H]
    \centering
        \begin{subfigure}{0.4\textwidth}
            \includegraphics[width=0.9\textwidth]{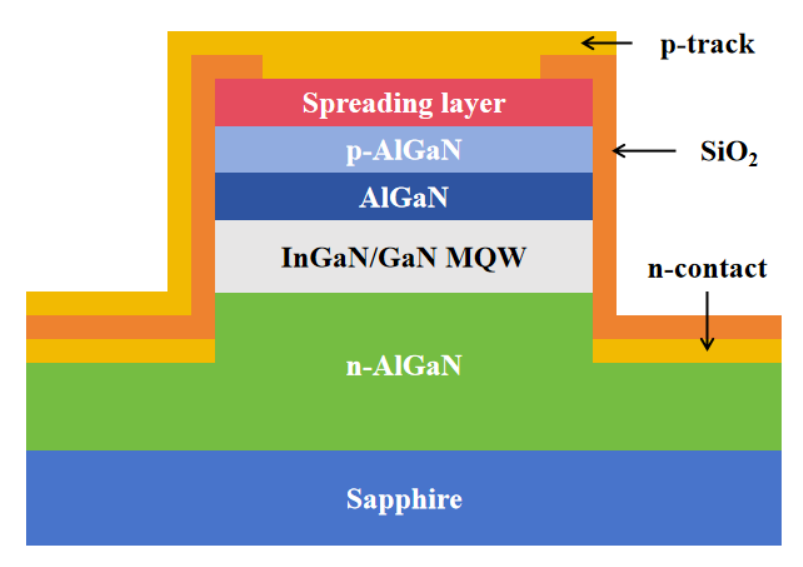}
            \caption{\label{mLEDsch}}
        \end{subfigure}
        \begin{subfigure}{0.4\textwidth}
            \includegraphics[width=0.9\textwidth]{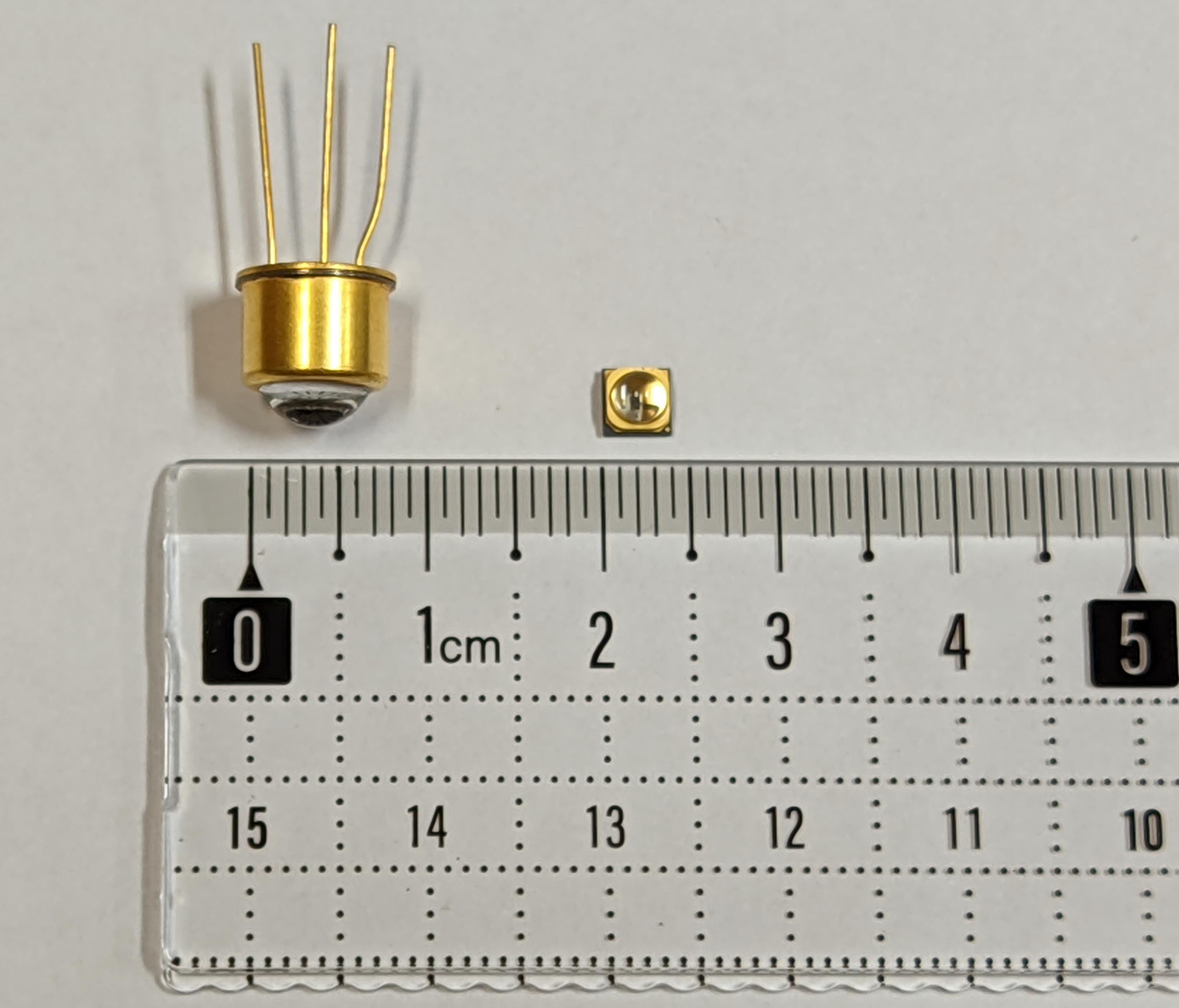}
            \caption{\label{mLED&cLED}}
        \end{subfigure}
    \caption{(\subref{mLEDsch}) Schematic structure of micro-LED, and (\subref{mLED&cLED}) photograph of 254 nm micro-LED and a common 255 nm LED in a TO-39 package with a lens front.}
    \label{mLED}
\end{figure}

By employing semiconductor technologies in manufacturing, the size of micro-LEDs is much smaller than that of common lighting LEDs, generally between 1-100 $\mu$m (see Figure \ref{mLEDsch})\cite{2GF, TDLA, MUMM}. Figure \ref{mLED&cLED} shows a micro-LED with a 254 nm peak wavelength and a full width at half maximum(FWHM) around 10 nm used in the experiment with a size of 100 $\mu$m next to a 255 nm LED in a TO-39 package with a lens front.

\begin{figure}[H]
    \centering
        \begin{subfigure}{0.4\textwidth}
            \includegraphics[width=0.9\textwidth]{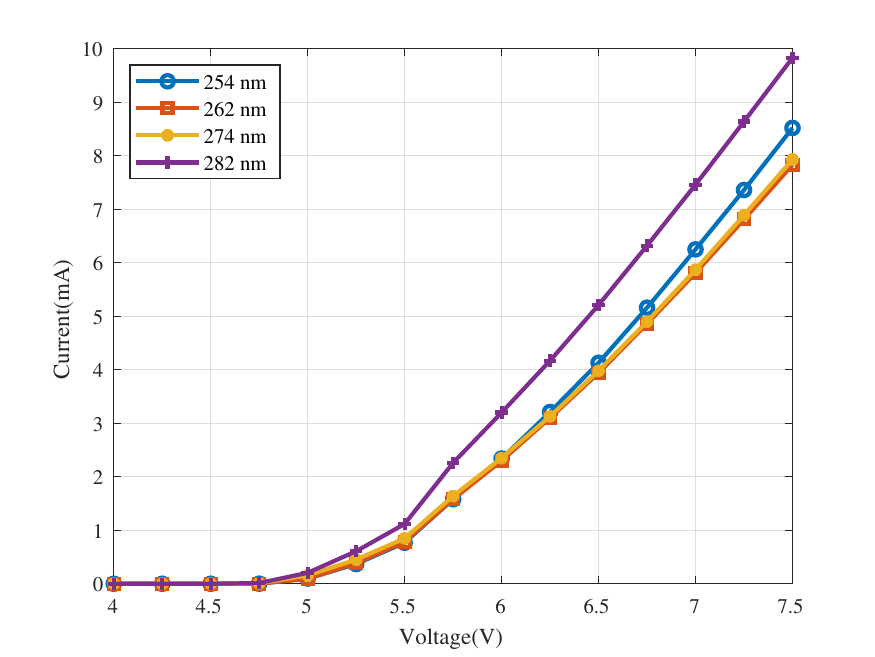}
            \caption{\label{UIP.UI}}
        \end{subfigure}
        \begin{subfigure}{0.4\textwidth}
            \includegraphics[width=0.9\textwidth]{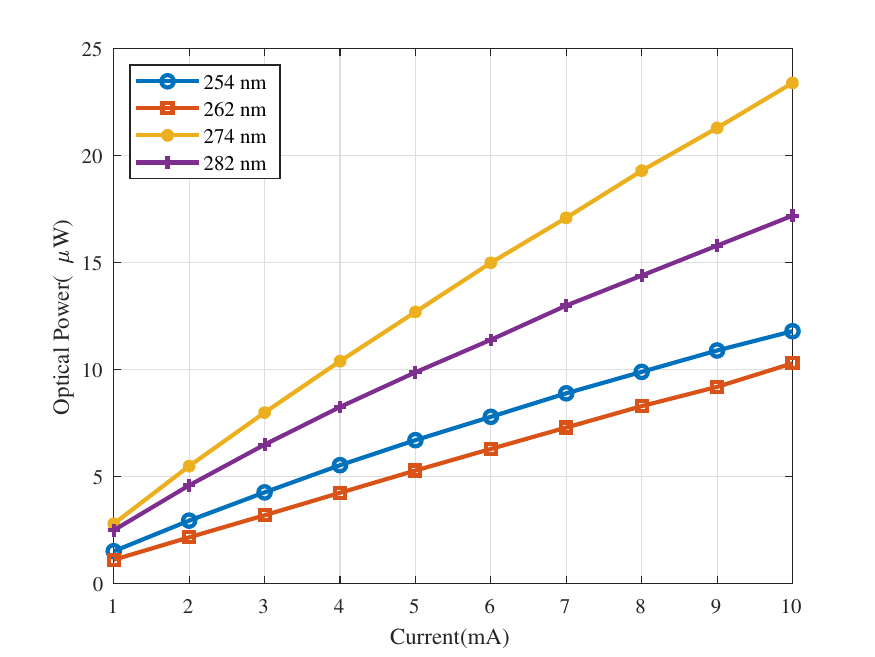}
            \caption{\label{UIP.IP}}
        \end{subfigure}
        \begin{subfigure}{0.4\textwidth}
            \includegraphics[width=0.9\textwidth]{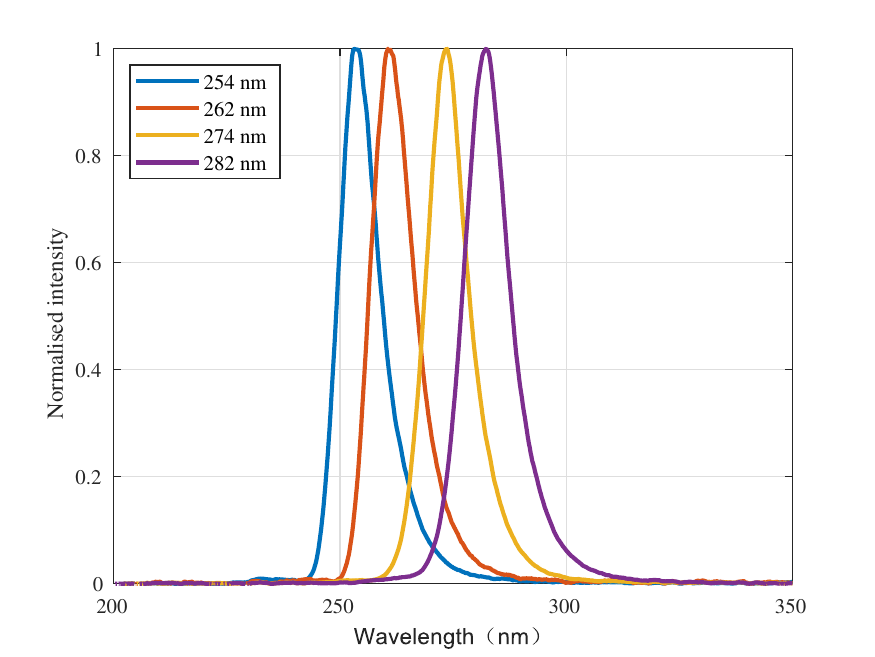}
            \caption{\label{UIP.Spe}}
        \end{subfigure}
   \caption{Characteristic performance plots of 4 micro-LEDs in the experiment. (\subref{UIP.UI}) Voltage (V) versus current (mA), (\subref{UIP.IP}) current (mA) versus optical power ($\mu$W), and (\subref{UIP.Spe}) measured emission spectrum.}
	\label{UIP}
\end{figure}

Representative curves showing the voltage vs current (V-I), current vs optical power (I-P), and emission spectrum of micro-LEDs with four different peak wavelengths are shown in Figure \ref{UIP}. In our measurement, the power meter was about 5 cm away from the micro-LEDs when testing the I-P curve.

\section{Photoelectrons emission and micro-LEDs}
\subsection{Experimental setup}

\begin{figure}[H]
    \centering
    \begin{subfigure}{0.4\textwidth}
        \includegraphics[width=0.9\textwidth]{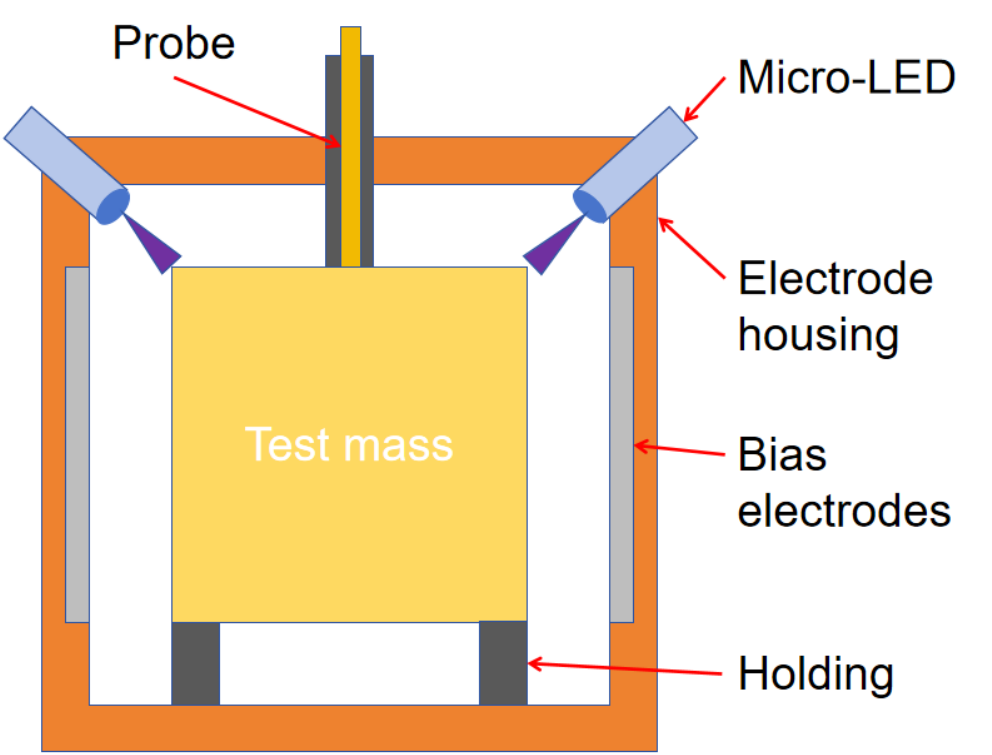}
        \caption{\label{sch&pho.sch}}
    \end{subfigure}
    \begin{subfigure}{0.3\textwidth}
        \includegraphics[width=0.9\textwidth]{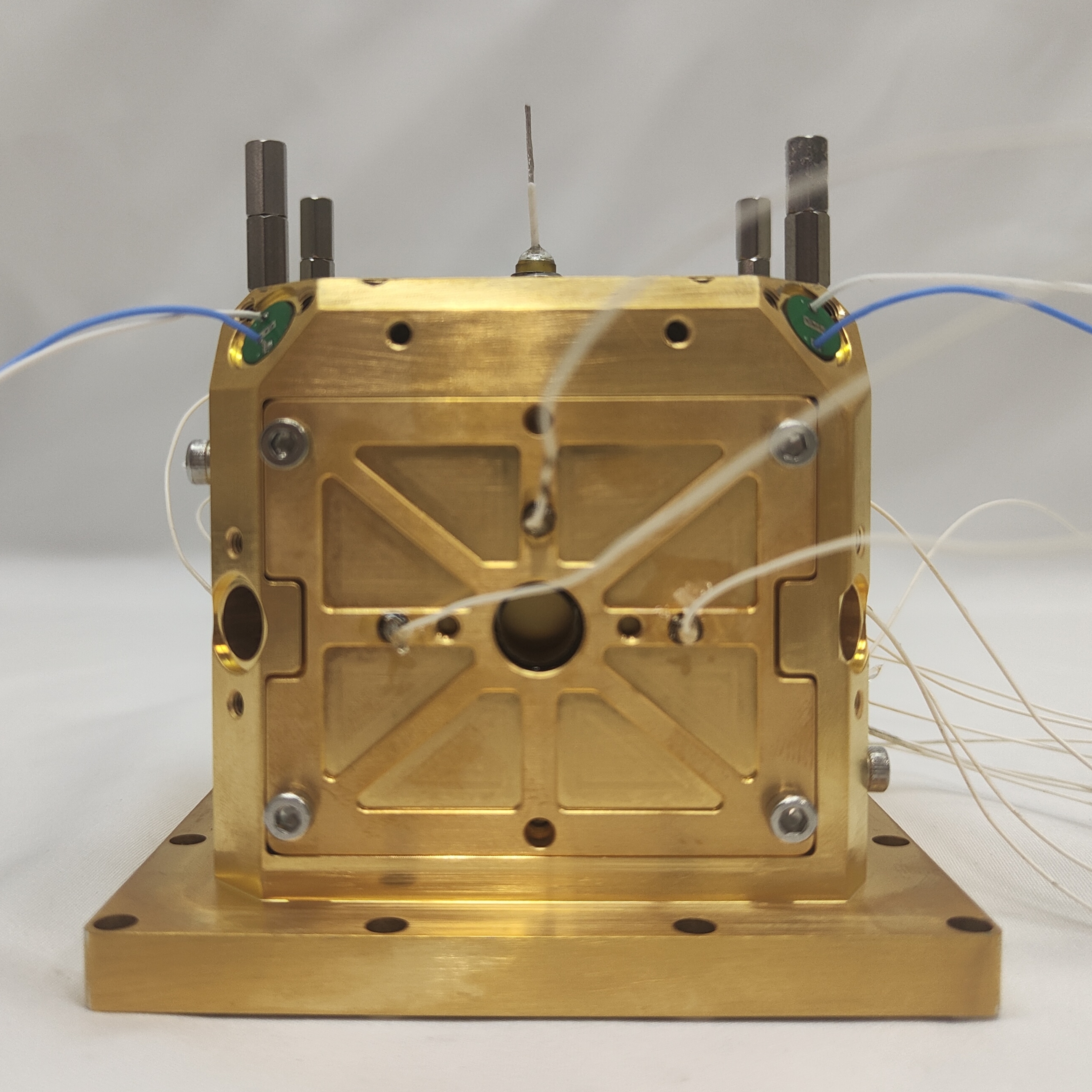}
        \caption{\label{sch&pho.pho}}
    \end{subfigure}
    \begin{subfigure}{0.3\textwidth}
        \includegraphics[width=0.9\textwidth]{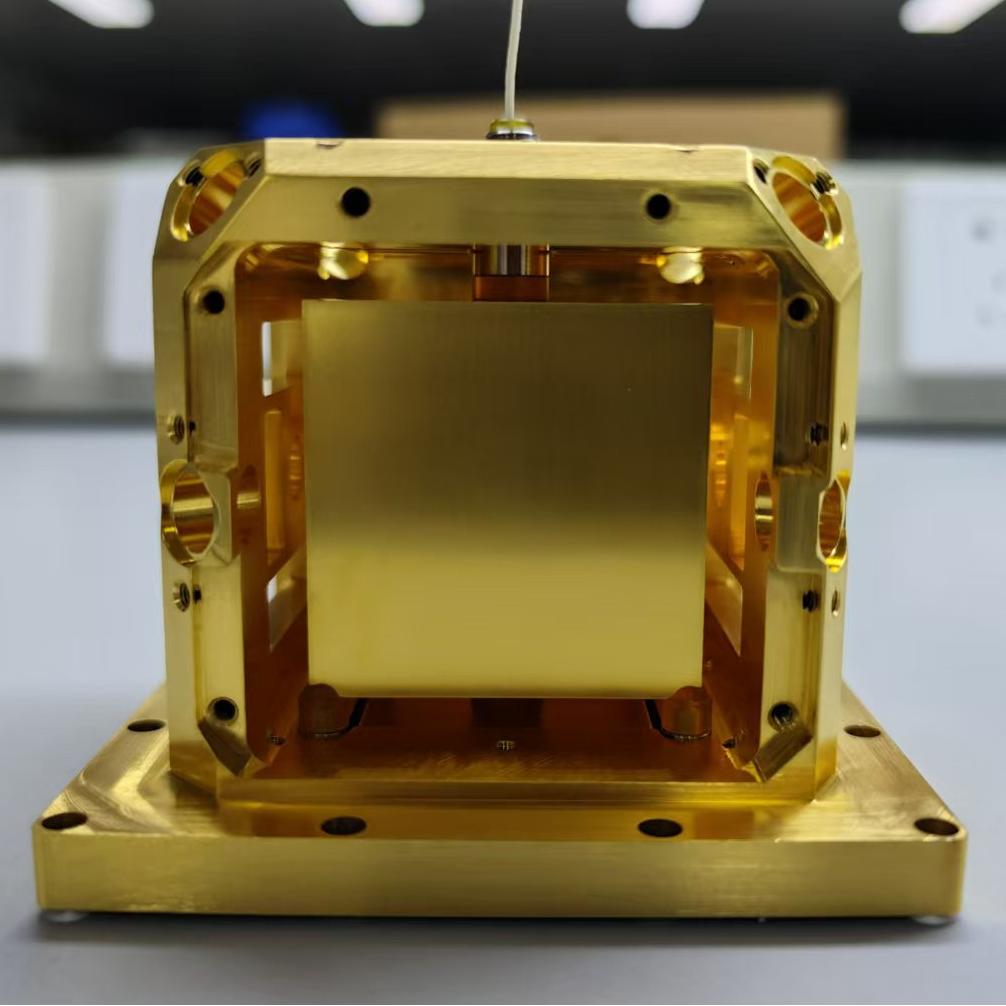}
        \caption{\label{sch&pho.pho2}}
    \end{subfigure}
    \caption{(\subref{sch&pho.sch}) Schematic of charge management experimental setup, (\subref{sch&pho.pho}) external of experimental setup showing electrode housing, and (\subref{sch&pho.pho2}) interior of experimental setup showing coated TM.}
    \label{sch&pho}
\end{figure}

Experiment was set up to demonstrate photoelectric effect generated by micro-LEDs(see Figure \ref{sch&pho}). The dimensions and electrodes distribution of the experimental setup for charge management system were designed with reference to LISA and LISA Pathfinder. This replaces our parallel plate experimental setup performed previously. A test mass(TM) was a cube with 46 mm in length and there was a gap of 5 mm between the test mass and the
electrodes mounted on the electrode housing. Both the surface of the test mass and the inner surface of the electrode housing were gold-coated with a thickness of 500 nm. The test mass was fixed inside the electrode housing by insulated Ultem-1000 holding tubes. Micro-LEDs were installed at the top corners of the housing and directly illuminate the test mass. A contact probe was mounted on the centre of the upper surface of the test mass to read out the potential of the test mass. Bias electrodes were grounded or connected to an external voltage source to provide the bias voltage.

\subsection{Test results}

\begin{figure}[H]
\centering
	\includegraphics[width=0.5\linewidth]{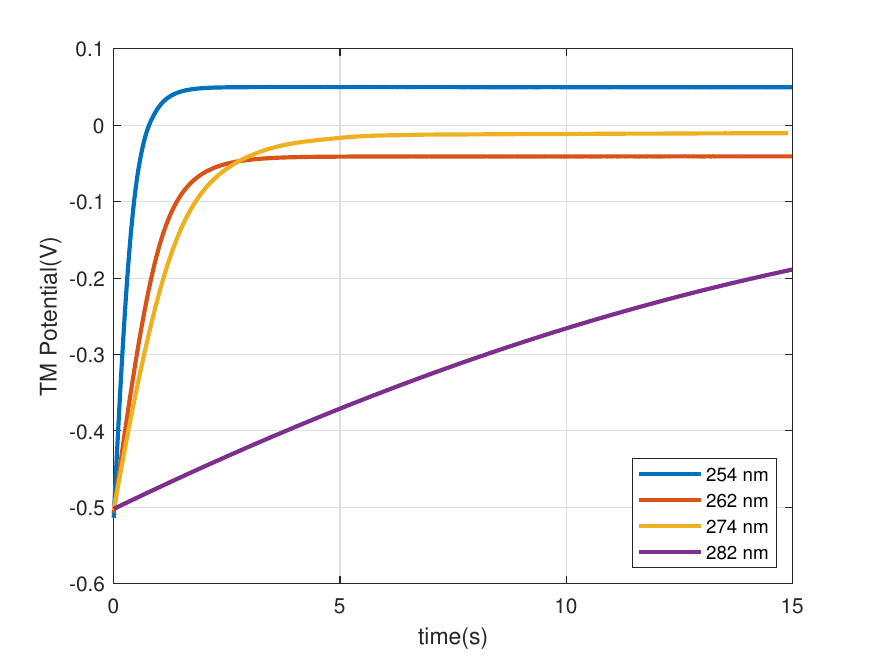}
	\caption{TM potential variations due to photoelectrons emitted by micro-LEDs with different wavelengths. Potential is measured relative to the electrode housing which is grounded and the drive current of micro-LED is 0.1 mA.}
	\label{multiemit}
\end{figure}

We present the results of several runs used to validate the functionality of the system. Firstly, four micro-LEDs with peak wavelengths of 254 nm, 262 nm, 274 nm and 282 nm were used for the charge management experiments. The drive current of micro-LEDs was set to 0.1mA, and the initial potential of the test mass($V_{\rm TM}$) was set to a negative potential with bias electrodes grounded. When the micro-LED was turned on, photoelectrons were emitted from test mass to electrode housing, which led to a reduction of charges on TM(Figure \ref{multiemit}).

\begin{figure}[H]
    \centering
	\begin{subfigure}{0.4\textwidth}   	
            \includegraphics[width=0.9\textwidth]{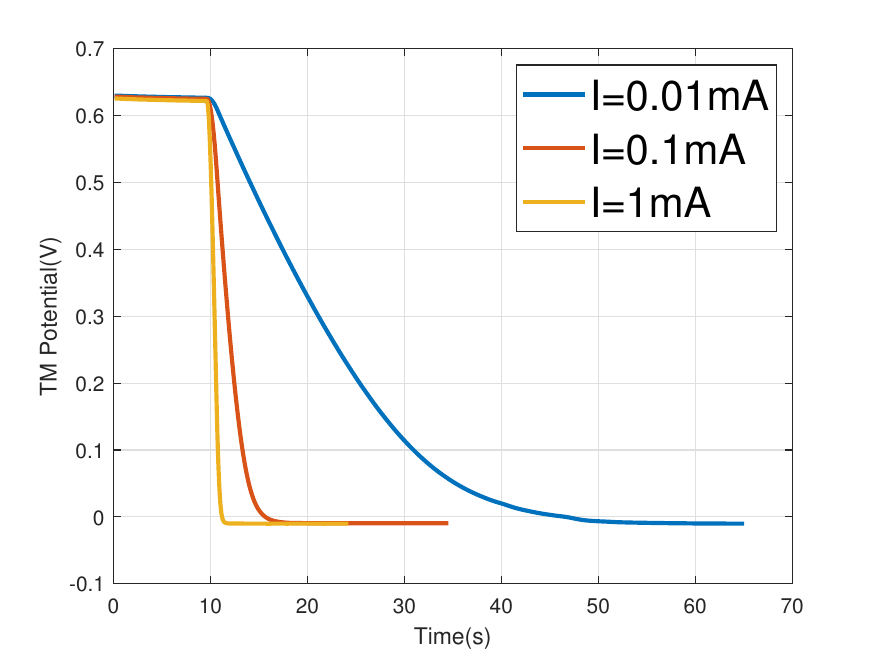}
            \caption{\label{275currentchange1}}
        \end{subfigure}
	\begin{subfigure}{0.4\textwidth}
    	\includegraphics[width=0.9\textwidth]{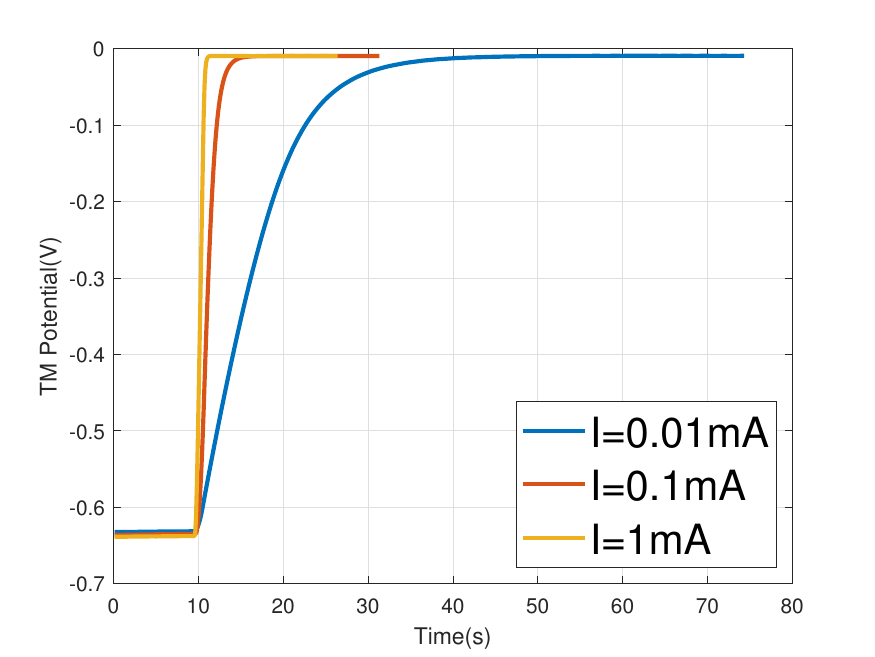}
            \caption{\label{275currentchange2}}
        \end{subfigure}
	\caption{The effect of the magnitude of the drive current on the system performance. (\subref{275currentchange1}) TM with positive initial voltage for 274 nm micro-LED at 0.01 mA, 0.1 mA, and 1 mA driving currents, and (\subref{275currentchange2}) TM with negative initial voltage for 274 nm micro-LED at 0.01 mA, 0.1 mA, and 1 mA driving currents.}
	\label{275currentchange}
\end{figure}

In order to verify the effect of the magnitude of the drive current(and thus of the output optical power that is proportional to it) on the system performance, we performed runs in which we monitored the $V_{\rm TM}$ while operating a 274 nm micro-LED at 0.01 mA, 0.1 mA, and 1 mA(Figure \ref{275currentchange}). The bias electrodes were grounded, and $V_{\rm TM}$ was set with inertail potentials about $\pm0.6$ V. It can be seen that the time taken to reach the equilibrium decreases with the increase of the drive current.

\begin{figure}[H]
    \centering
	\begin{subfigure}{0.4\textwidth}
    	\includegraphics[width=0.9\textwidth]{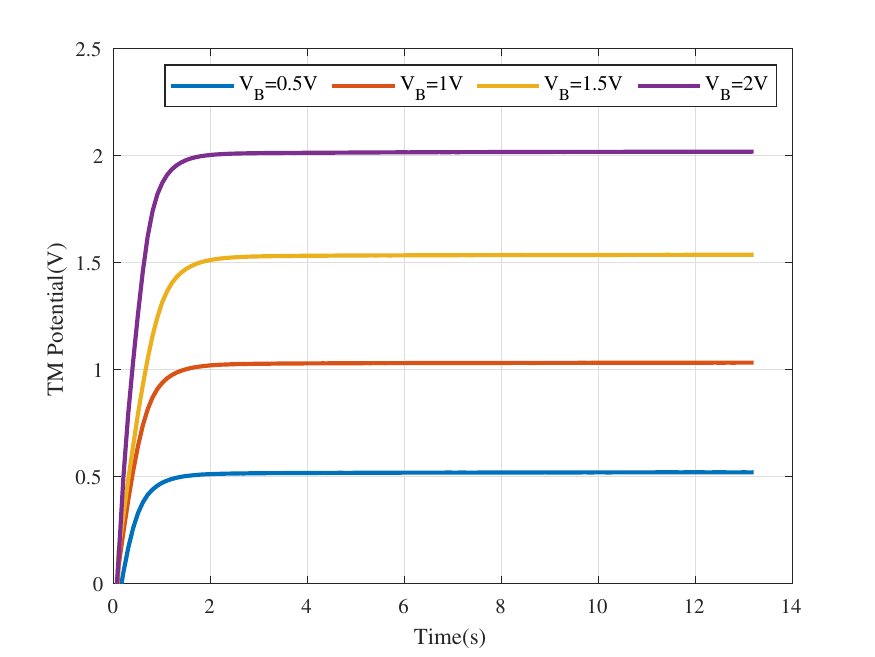}
            \caption{\label{biasp}}
        \end{subfigure}
        \begin{subfigure}{0.4\textwidth}
            \includegraphics[width=0.9\textwidth]{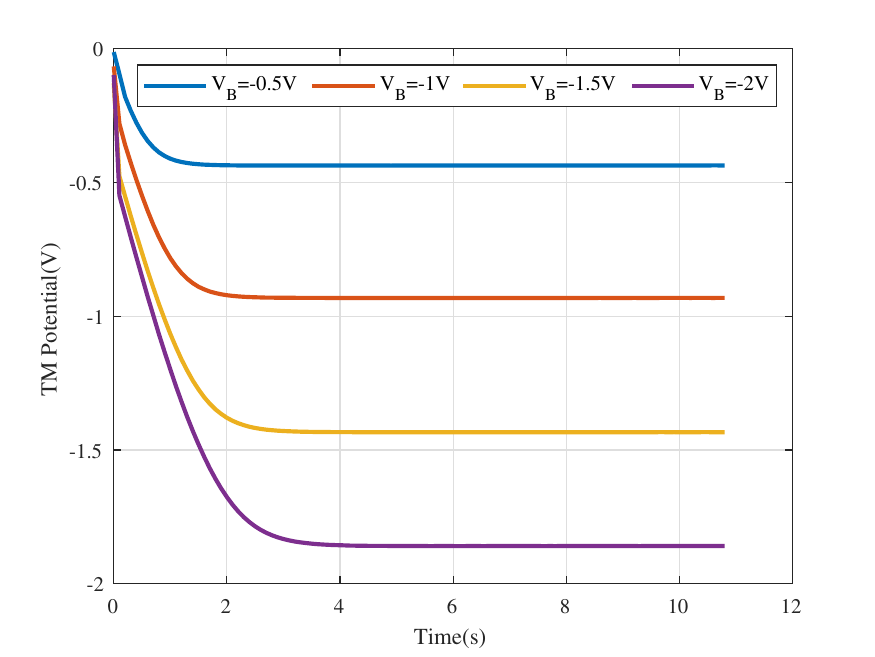}
            \caption{\label{biasm}}
        \end{subfigure}
        \begin{subfigure}{0.4\textwidth}
    	\includegraphics[width=0.9\textwidth]{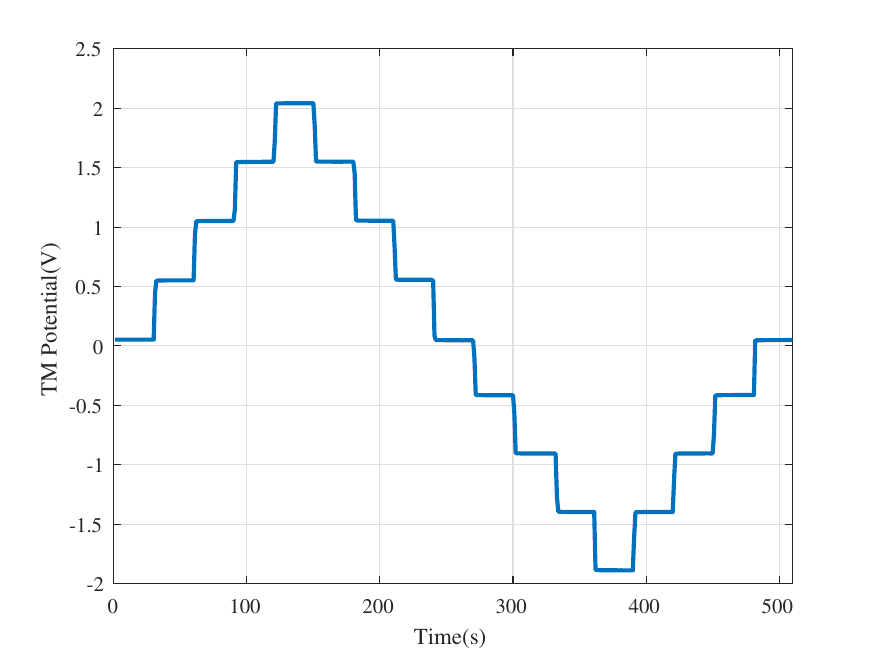}
            \caption{\label{bias1}}
        \end{subfigure}
        \begin{subfigure}{0.4\textwidth}
    	\includegraphics[width=0.9\textwidth]{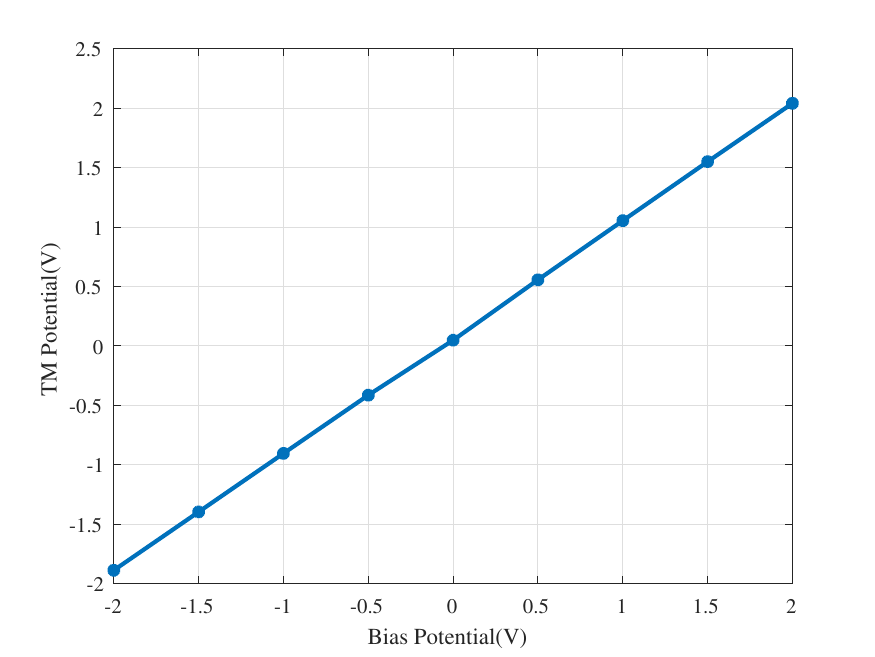}
            \caption{\label{bias2}}
        \end{subfigure}
	\caption{Demonstration of the variation of $V_{\rm TM}$ affected by $V_{\rm B}$. (\subref{biasp}) $V_{\rm TM}$ varies with positive $V_{\rm B}$ from +0.5 V to +2 V, (\subref{biasm}) $V_{\rm TM}$ varies with negative $V_{\rm B}$ from -0.5 V to -2 V, (\subref{bias1}) $V_{\rm TM}$ as a function of time with $V_{\rm B}$ varied from 0 V to 2 V, to -2 V and back to 0 V in steps of 0.5 V, and (\subref{bias2}) $V_{\rm TM}$ as a function of $V_{\rm B}$.}
	\label{bias}
\end{figure}

The potential of the bias electrode($V_{\rm B}$) will affect the potential variation of TM. Figure \ref{bias} shows a verification of the $V_{\rm B}$ varies for neutralization of TM potentials of $\pm0.5$ V, $\pm1$ V, $\pm1.5$ V, and $\pm2$ V. 

In this test, we used a 254 nm micro-LED where the drive current was set to 1 mA, and the potential of the bias electrode was set from +0.5 V to +2 V(Figure \ref{biasp}), and -0.5 V to -2 V(Figure \ref{biasm}). It can be observed that the rate of charge variation increases with increasing potential of the bias electrode.  A continuous test was conducted where the bias electrodes varied from 0 V to 2 V, to -2 V and back to 0 V in steps of 0.5 V, and each step lasted about 30 seconds. while TM potential was monitored until it reached equilibrium(Figure \ref{bias1}). The $V_{\rm TM}$ equilibrium potentials for the nine $V_{\rm B}$ settings are shown in Figure \ref{bias2} , and the fit for the data is:

\begin{equation} \label{eq:fit1}
{V_{\rm TM}(\rm V)} = 0.976 \times {V_{\rm B}(\rm V) + 0.055(\rm V)}
\end{equation}

\section{Micro-LED and continuous charge control}

During the ascending and descending phase of a solar cycle, solar activity is relatively quiet and the positively charged particles are dominated by galactic cosmic rays. The flux of these particles entering a spacecraft may be well modeled. In these scenarios, continuous charge control strategy may be employed in the sense that parameters for the open loop charge management, like for instance, optical power, may be fine tuned in such a way the voltage (residual charges) on the test mass may be kept to a very stable and low level, in order to meet the requirement for gravitational wave detection.  

The output optical power of LED can be adjusted by adjusting the drive current within a certain range or duty cycle of PWM. Stanford University has successfully flown a technology demonstration of AC charge management with UV-LEDs by adjusting the drive current(0-10 mA) and duty cycle(0-100{\%})\cite{GTF}. UV-LED with drive current of 17.5 mA and duty cycle of 6{\%} may reach a low optical power of approximately 200 pW\cite{ACM}. 

In order to achieve precise light power output needed to further reduce the drive current or duty cycle, which puts a stringent demand on the LED's precise control and long-term stability. Due to the smaller size, micro-LEDs not only possess a higher modulation bandwidth(to GHz), but also offer finer current adjustment capabilities(to nA), enabling more precise control of the optical output power\cite{MDT}. These advantages make micro-LEDs a promising alternative scheme for charge management system.

\subsection{Experimental results.}

\begin{figure}[H]
    \centering
        \begin{subfigure}{0.4\textwidth}
    	\includegraphics[width=0.9\textwidth]{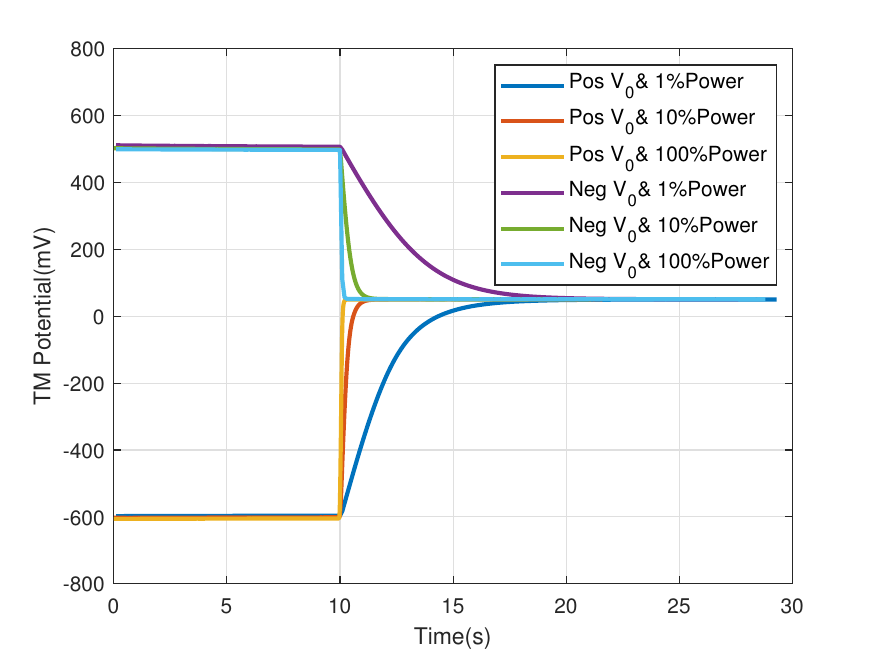}
            \caption{\label{CurDutChange.255}}
        \end{subfigure}
         \begin{subfigure}{0.4\textwidth}
    	\includegraphics[width=0.9\textwidth]{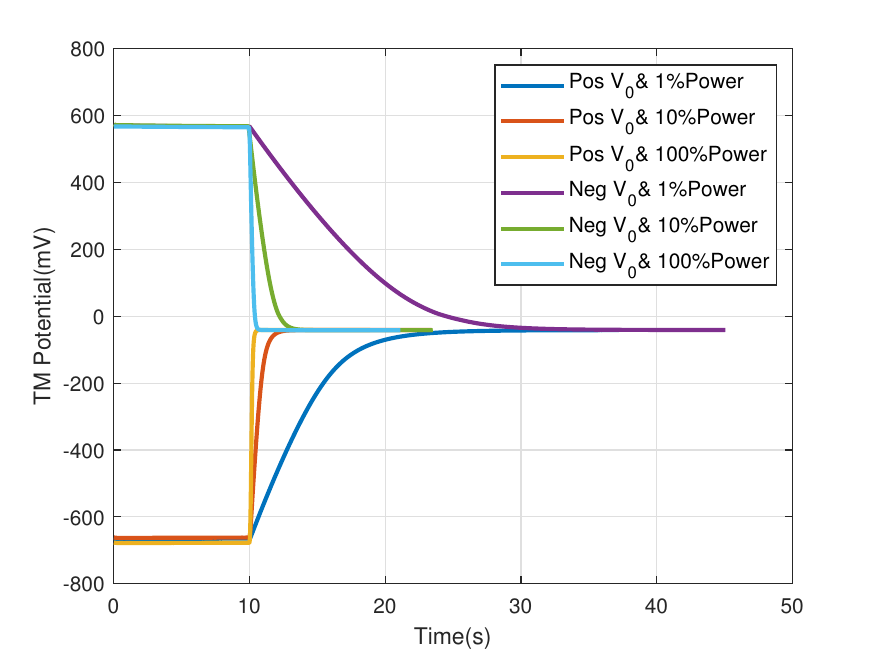}
            \caption{\label{CurDutChange.265}}
        \end{subfigure}
         \begin{subfigure}{0.4\textwidth}
    	\includegraphics[width=0.9\textwidth]{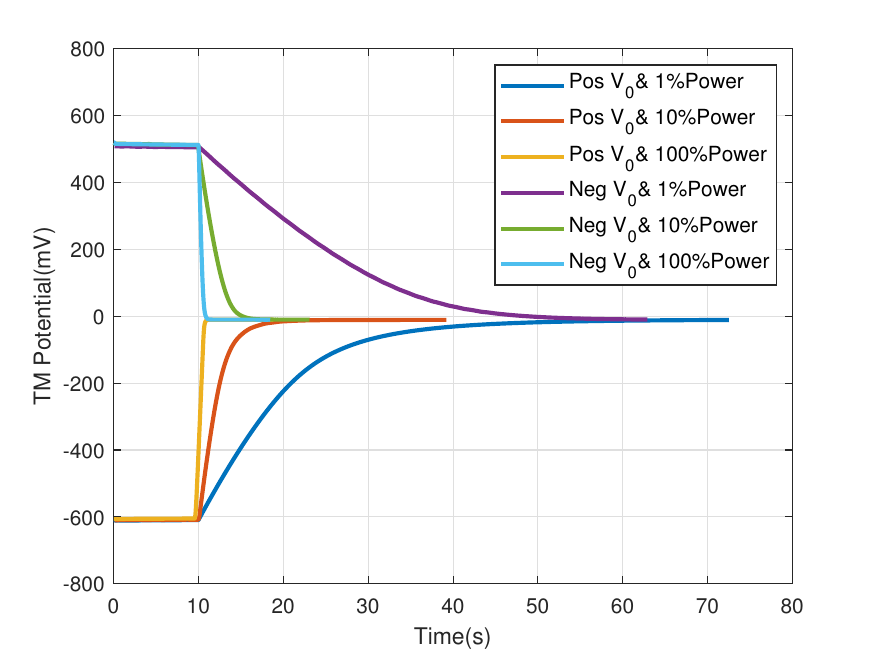}
            \caption{\label{CurDutChange.275}}
        \end{subfigure}
         \begin{subfigure}{0.4\textwidth}
    	\includegraphics[width=0.9\textwidth]{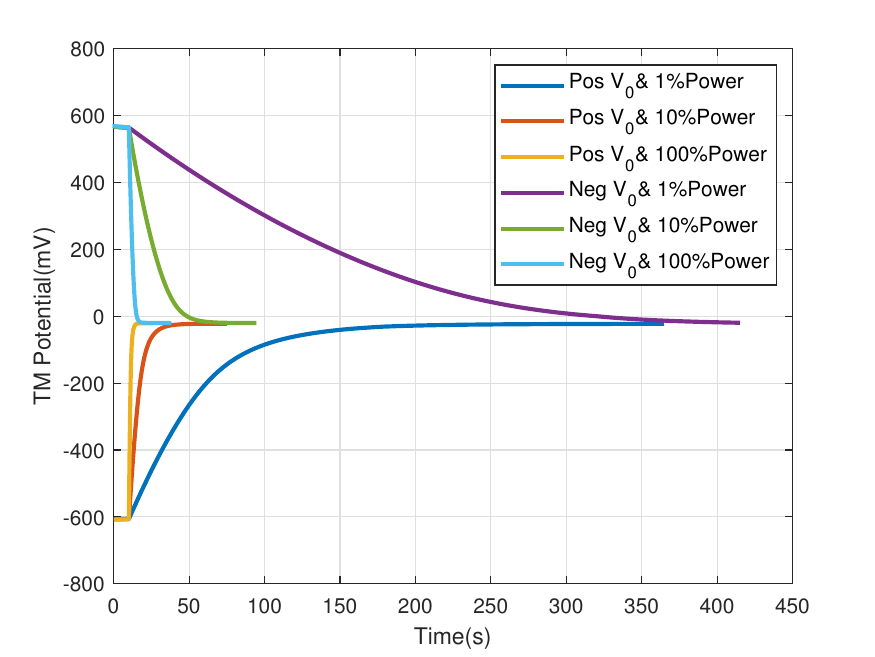}
            \caption{\label{CurDutChange.285}}
        \end{subfigure}
    
	\caption{Test results showing the measured TM potentials when setting different initial potentials of TM and adjusting the optical power of four different wavelengths of micro-LEDs. No bias electrode voltage is applied. (\subref{CurDutChange.255}) 254 nm, (\subref{CurDutChange.265}) 262 nm, (\subref{CurDutChange.275}) 274 nm,  and (\subref{CurDutChange.285}) 282 nm.}
	\label{CurDutChange}
\end{figure}

The above four micro-LEDs were selected for the charge management experiments. By setting the drive current to 0.1 mA and 1 mA and modulating at 1 kHz with the duty cycles to 10{\%}, 50{\%} and 100{\%}, the relative intensities of the output optical power($P_{\rm r}$) of the micro-LEDs were controlled to be 1{\%}, 5{\%}, 10{\%}, 50{\%} and 100{\%}, respectively. The potential change of the test mass was monitored throughout the process.  Figure \ref{CurDutChange} shows the results when setting different initial voltages of test mass and adjusting $P_{\rm r}$ of the four micro-LEDs to 1{\%}, 10{\%} and 100{\%}.

Firstly, an initial positive or negative potential($V_{\rm 0}$) was given to the test mass, due to the high resistance between TM and surroundings, the potential of the test mass remained nearly unchanged until the micro-LED was switched on at about 10th seconds. Once the micro-LED was turned on, the TM potential rapidly approached to the equilibrium potential($V_{\rm eq}$) regardless of whether $V_{\rm 0}$ was greater or less than $V_{\rm eq}$.

For micro-LED with a given wavelength, like 254 nm in Fig \ref{CurDutChange.255}, the change of output optical power generates a variation of the incident photon flux, and the time required to reach equilibrium potential inversely proportional to the increase of photon flux while the equilibrium voltage is independent of it. Similar results were obtained for other wavelengths as shown in Fig \ref{CurDutChange.265} - Fig \ref{CurDutChange.285}. For micro-LEDs of different wavelengths, the equilibrium potentials were different. Several wavelengths of micro-LEDs used in the experiment can control the potential of the test mass to the range of 100 mV, with the equilibrium voltage of 274 nm wavelength micro-LED being closest to zero. Table \ref{Equilibrium} gives the equilibrium voltage at different wavelengths.

\begin{table}[H]
    \centering
    \caption{Equilibrium potentials with different wavelength micro-LEDs.}
    \begin{tabular}{cc}
    \hline
    Wavelength(nm) & Equilibrium potential(mV)\\
    \hline
    254 & 50\\
    262 & -41\\
    274 & -10\\
    282 & -21\\
    \hline
    \end{tabular}
    \label{Equilibrium}
\end{table}

\begin{figure}[H]
    \centering
        \begin{subfigure}{0.4\textwidth}
    	\includegraphics[width=0.9\textwidth]{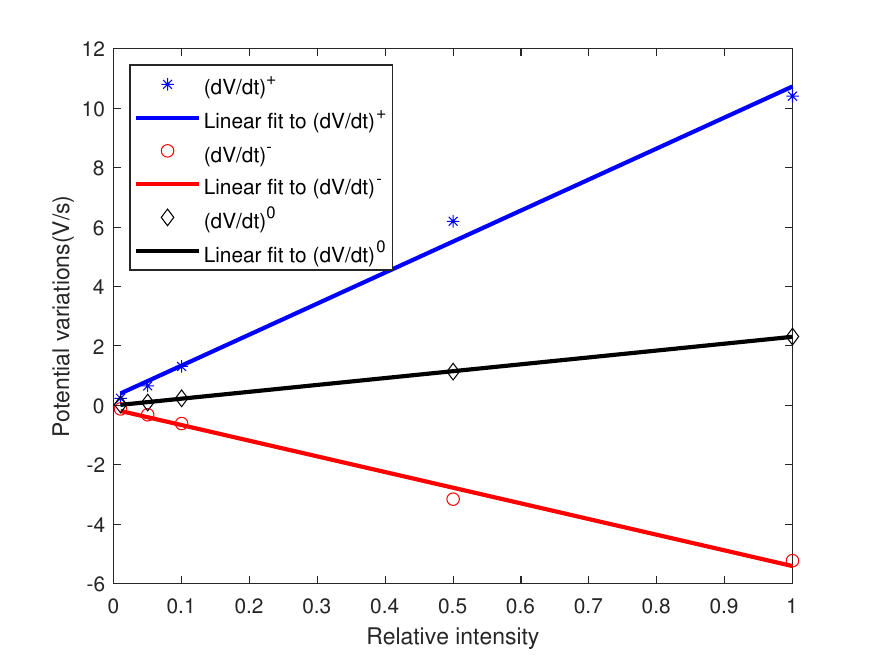}
            \caption{\label{dVdt.255}}
        \end{subfigure}
        \begin{subfigure}{0.4\textwidth}
    	\includegraphics[width=0.9\textwidth]{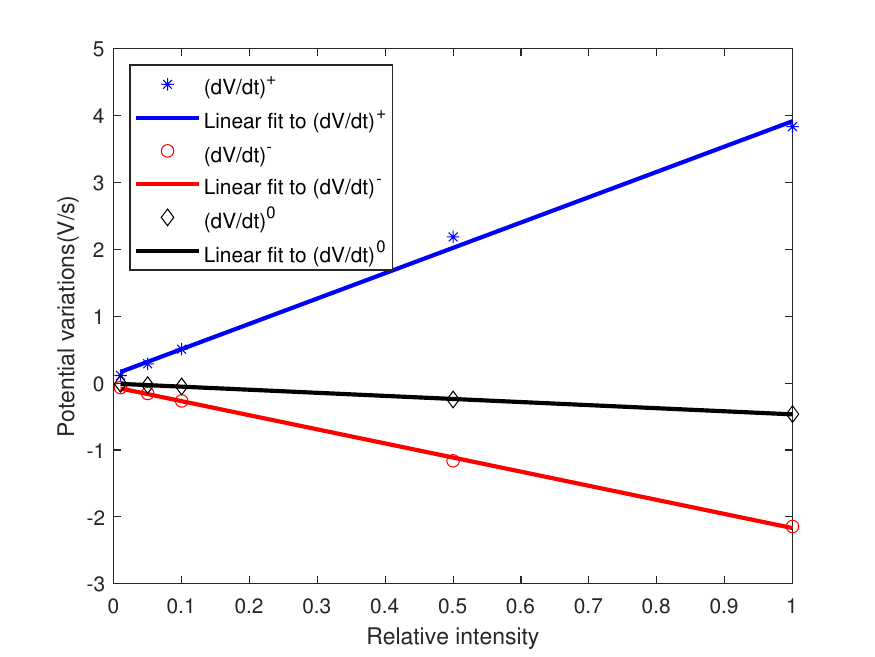}
            \caption{\label{dVdt.265}}
        \end{subfigure}
        \begin{subfigure}{0.4\textwidth}
    	\includegraphics[width=0.9\textwidth]{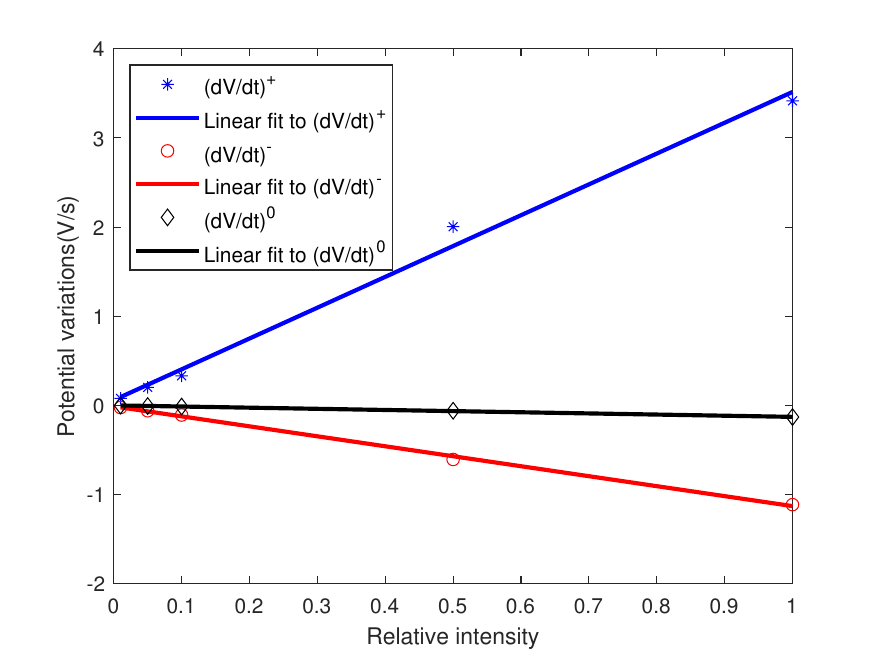}
            \caption{\label{dVdt.275}}
        \end{subfigure}
        \begin{subfigure}{0.4\textwidth}
    	\includegraphics[width=0.9\textwidth]{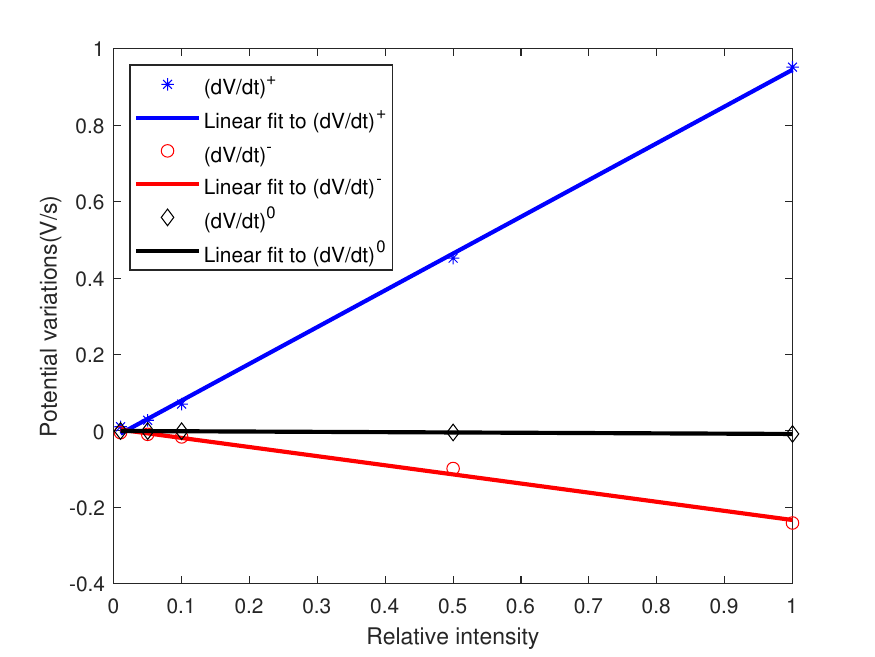}
            \caption{\label{dVdt.285}}
        \end{subfigure}
	\caption{Plots showing the measured rates of potential variations of TM at the start of initial positive potential, the start of initial negative potential and $V_{\rm TM}=0$ for the four wavelength micro-LEDs. (\subref{dVdt.255}) 254 nm, (\subref{dVdt.265}) 262 nm, (\subref{dVdt.275}) 274 nm, and (\subref{dVdt.285}) 282 nm.}
	\label{dVdt}
\end{figure}

The potential variations of test mass will increase with increasing the output optical power. Figure \ref{dVdt} shows the maxima of the potential variations for the five different output optical powers at the start of initial positive potential$(({\rm d}V/{\rm d}t)^+)$, and at the start of the negative potential$(({\rm d}V/{\rm d}t)^-)$. These are used to measure the maximum discharge rate achievable during charge control after the test mass has accumulated a significant amount of charge (such as after its release). Additionally, the rate of potential variations near 0V($({\rm d}V/{\rm d}t)^0$) is used to verify the enhancement of fine charge control capability by adjusting the driving current or PWM duty cycle of the micro-LED to improve optical power resolution.

It may be seen that the rate of potential variations increases linearly as the output optical power increases, i.e., the rate of potential variations can be changed by changing the drive current or duty cycle. When the drive current was set to 1 mA and the duty cycle was set to 100{\%}, the maxima of the potential variations of TM were more than 1 V/s for all four wavelengths micro-LEDs, and therefore, the potential of the TM can be controlled within 100 mV in less than 1 second. And under the same intensity, the short wavelength micro-LED had a faster rate of potential variations. And when the optical power was reduced, such as reducing the drive current(from 1 mA to 0.1 mA) or reducing the duty cycle(from 100{\%} to 10{\%}), the rate of potential variations of the test mass was also reduced. 

Considering the performance of micro-LEDs under continuous mode first, we can obtain the relationships between the rate of potential variations near 0 V($({\rm d}V/{\rm d}t)^0$) with the relative intensities of the output optical power($P_r$). The fits for the data are:

\begin{equation} \label{eq:fit254}
({\rm d}V/{\rm d}t)^0(254\ {\rm nm})({\rm V/s}) = 2.31 \times P_r(254\ \rm nm)
\end{equation}
\begin{equation}\label{eq:fit262}
({\rm d}V/{\rm d}t)^0(262\ {\rm nm})({\rm V/s}) = 0.46 \times P_r(262\ \rm nm)
\end{equation}
\begin{equation}\label{eq:fit274}
({\rm d}V/{\rm d}t)^0(274\ {\rm nm})({\rm V/s}) = 0.13 \times P_r(274\ \rm nm)
\end{equation}
\begin{equation}\label{eq:fit282}
({\rm d}V/{\rm d}t)^0(282\ {\rm nm})({\rm V/s}) = 0.008 \times P_r(282\ \rm nm)
\end{equation}

It may be seen that $({\rm d}V/{\rm d}t)^0$ can be reduced by decreasing $P_r$ through adjusting the drive current and duty cycle. The drive current of micro-LED can be in the order of $\mu$A or even nA, , and micro-LEDs are able to generate subnanosecond light pulses output. Therefore, it can be expected that when the drive current is set to 1 $\mu$A and the pulse width is set to 1 $\mu$s, the rate of variation of charges is expected to be reduced to an order of 10 charges/s, given the capacitance of TM to ground is of the order of 10 pF. It meets the continuous charge management requirement of when solar activities are quiet, as often the case in the ascending and descending phase of a solar cycle.

Charge management systems usually employ optical fibre to transport the light into electrode housing. Optical fibre will have attenuation, and the degree of attenuation will increase with exposure to UV light. In the worst scenario when the fibre attenuation is 1{\%}, the potential variation rate will attenuate from $>$ 1 V/s to $>$ 10 mV/s, while the corresponding discharge rate is greater than $10^6$ charges/s, as required when solar energetic particle(SEP) event occurs. Higher discharge rate can be provided by either increasing the drive current or the number of micro-LEDs.

\begin{figure}[H]
    \centering
	\includegraphics[width=0.5\linewidth]{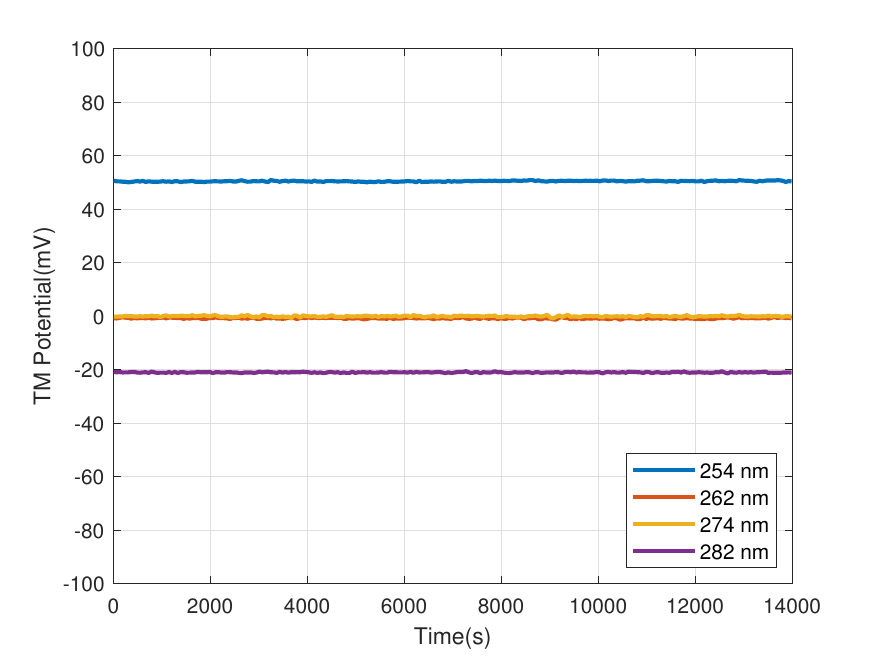}
	\caption{TM potential drift about 4 hours for 254 nm, 262 nm, 274 nm, and 282 nm micro-LEDs with drive current of 1mA.}
	\label{long}
\end{figure}

To determine the stability of charge management systems by using the micro-LED, observations were performed about 4 hours with the four micro-LEDs giving drive current of 1mA. During this period, the potential of TM was simultaneously monitored with the bias electrode grounded. Fig \ref{long} demostrates the potential drifts for the four micro-LEDs.

It can be seen that the TM potential remains stable within the target range of ±100 mV over relatively long periods of time and the variation is less than 1 mV during the 4 hours test. These experiments demonstrate that UV micro-LED and UV LED are in equal footing in continuous charge management.

\section{Space Qualification of micro-LED -- a preliminary study}
\subsection{Experimental setup}

\begin{figure}[H]
    \centering
        \begin{subfigure}{0.4\textwidth}
    	\includegraphics[width=0.9\textwidth]{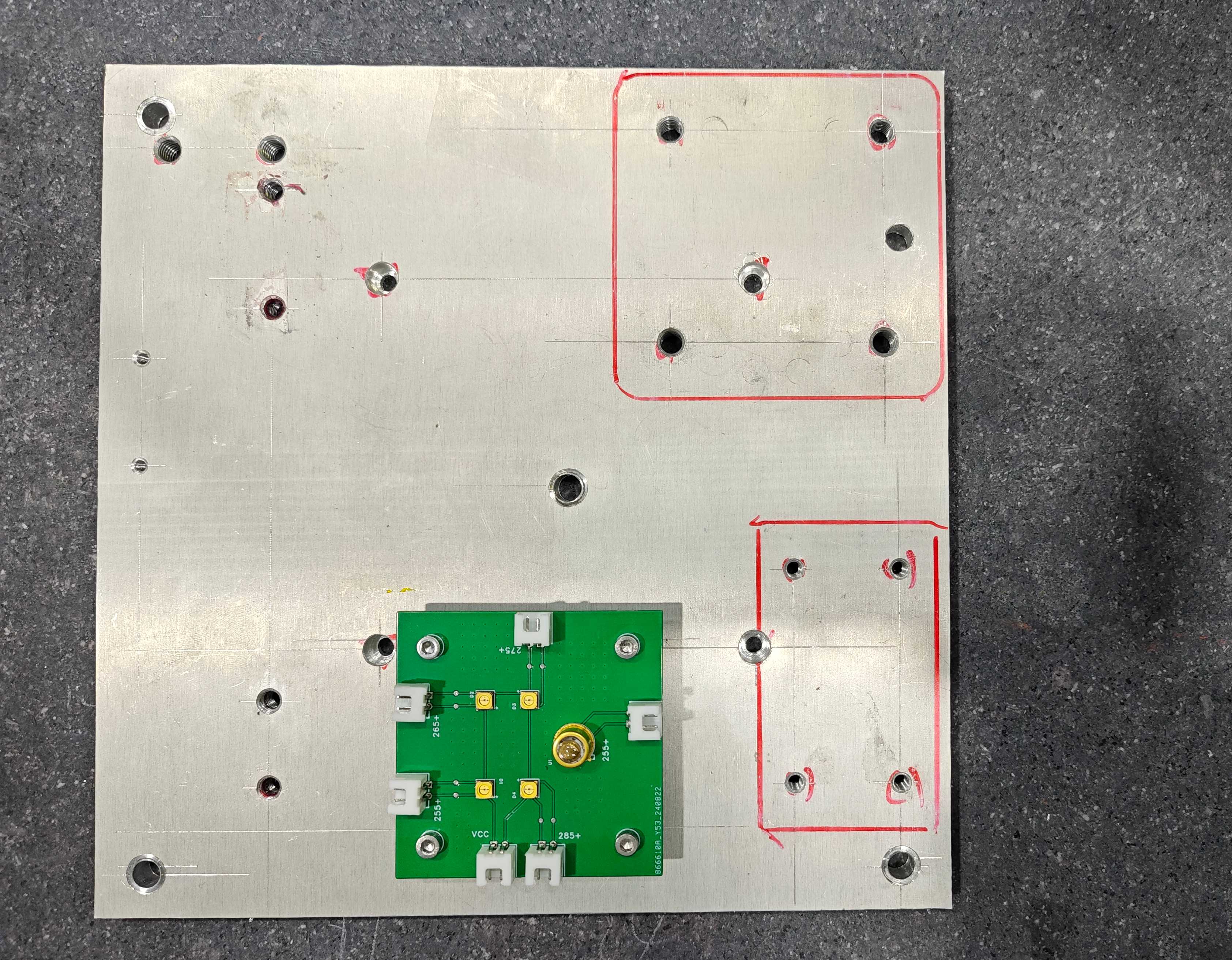}
            \caption{\label{setting}}
        \end{subfigure}
        \begin{subfigure}{0.4\textwidth}
    	\includegraphics[width=0.7\textwidth, angle=90]{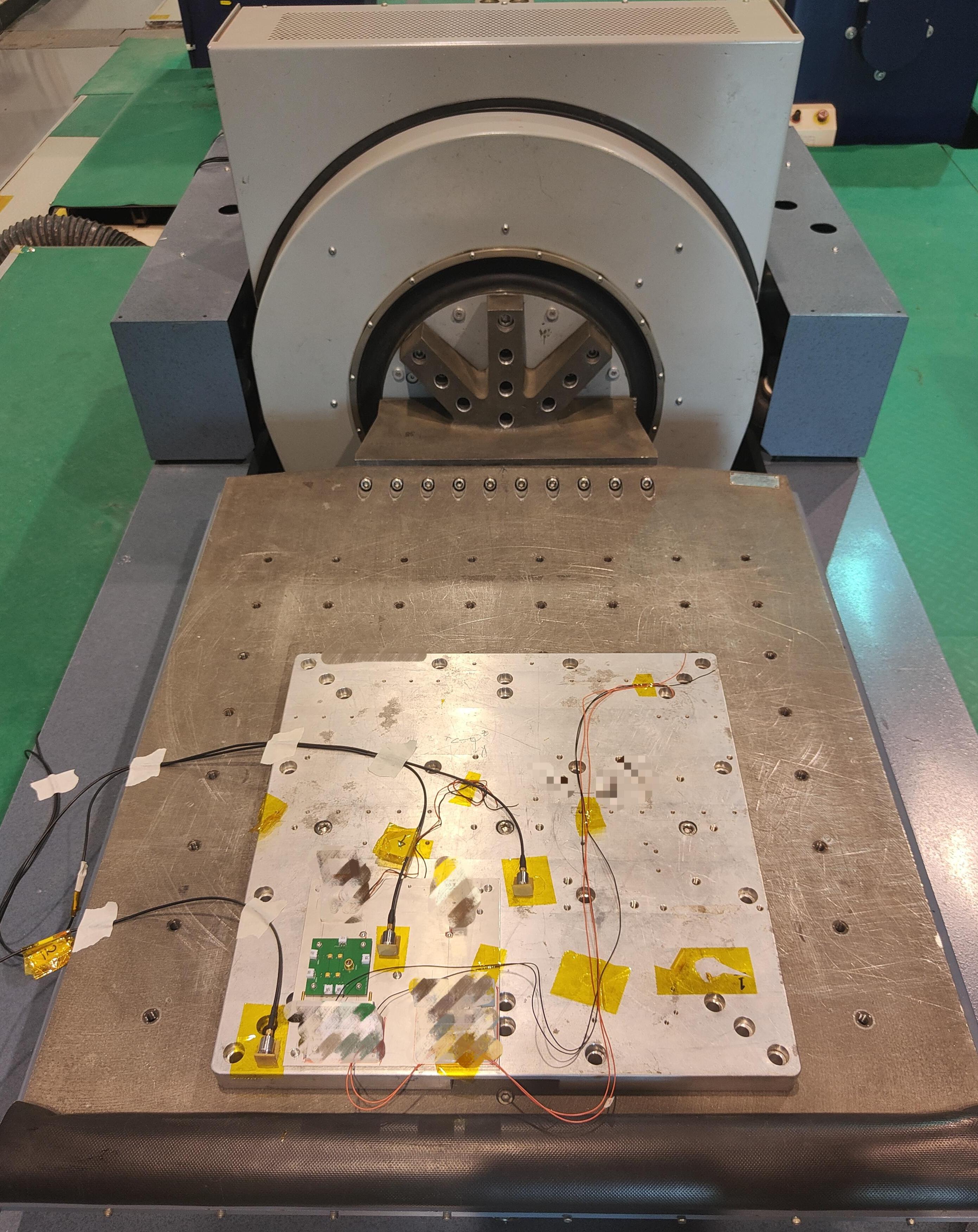}
            \caption{\label{vibration}}
        \end{subfigure}
        \begin{subfigure}{0.4\textwidth}
    	\includegraphics[width=0.9\textwidth]{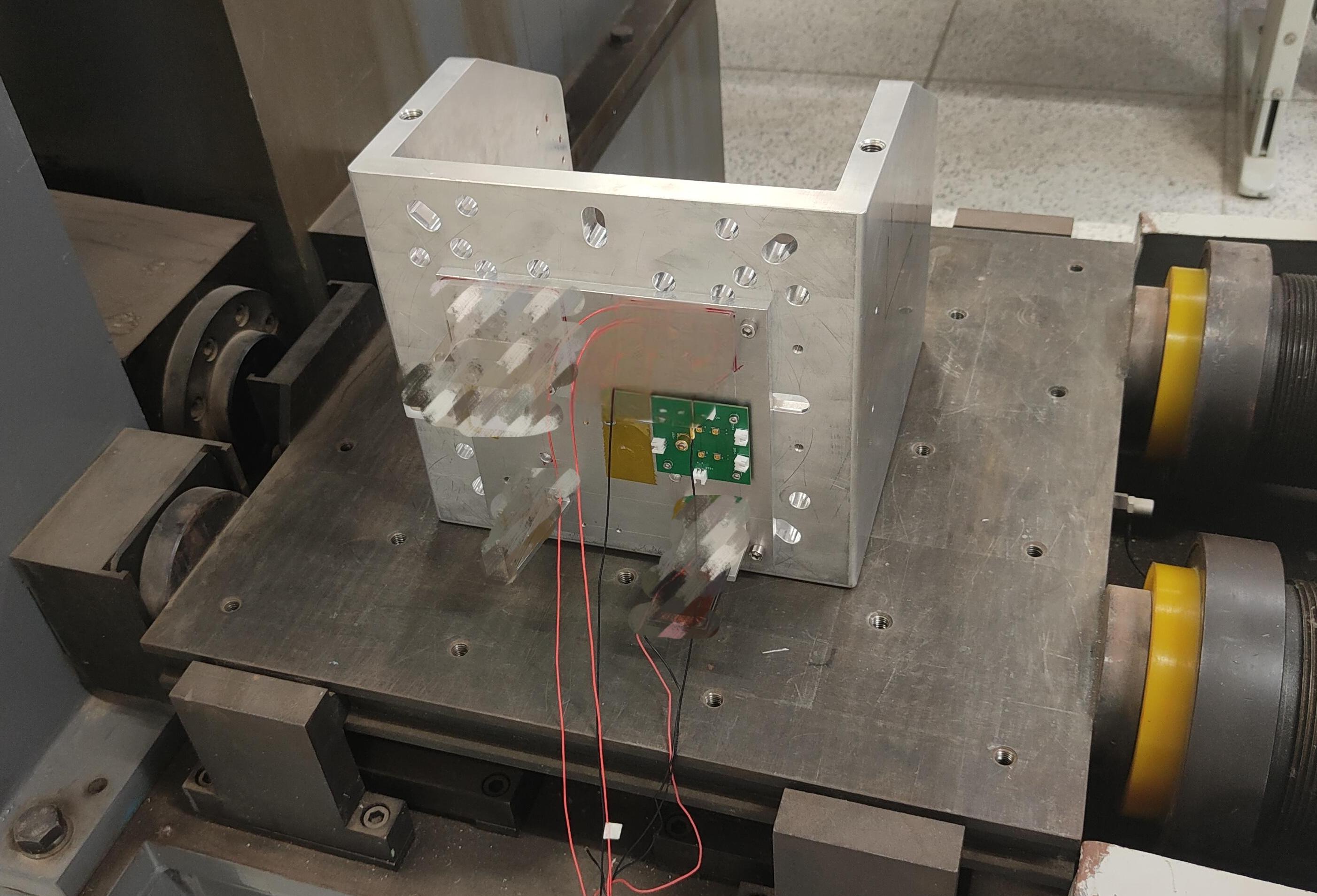}
            \caption{\label{shock}}
        \end{subfigure}
        \begin{subfigure}{0.4\textwidth}
    	\includegraphics[width=0.85\textwidth]{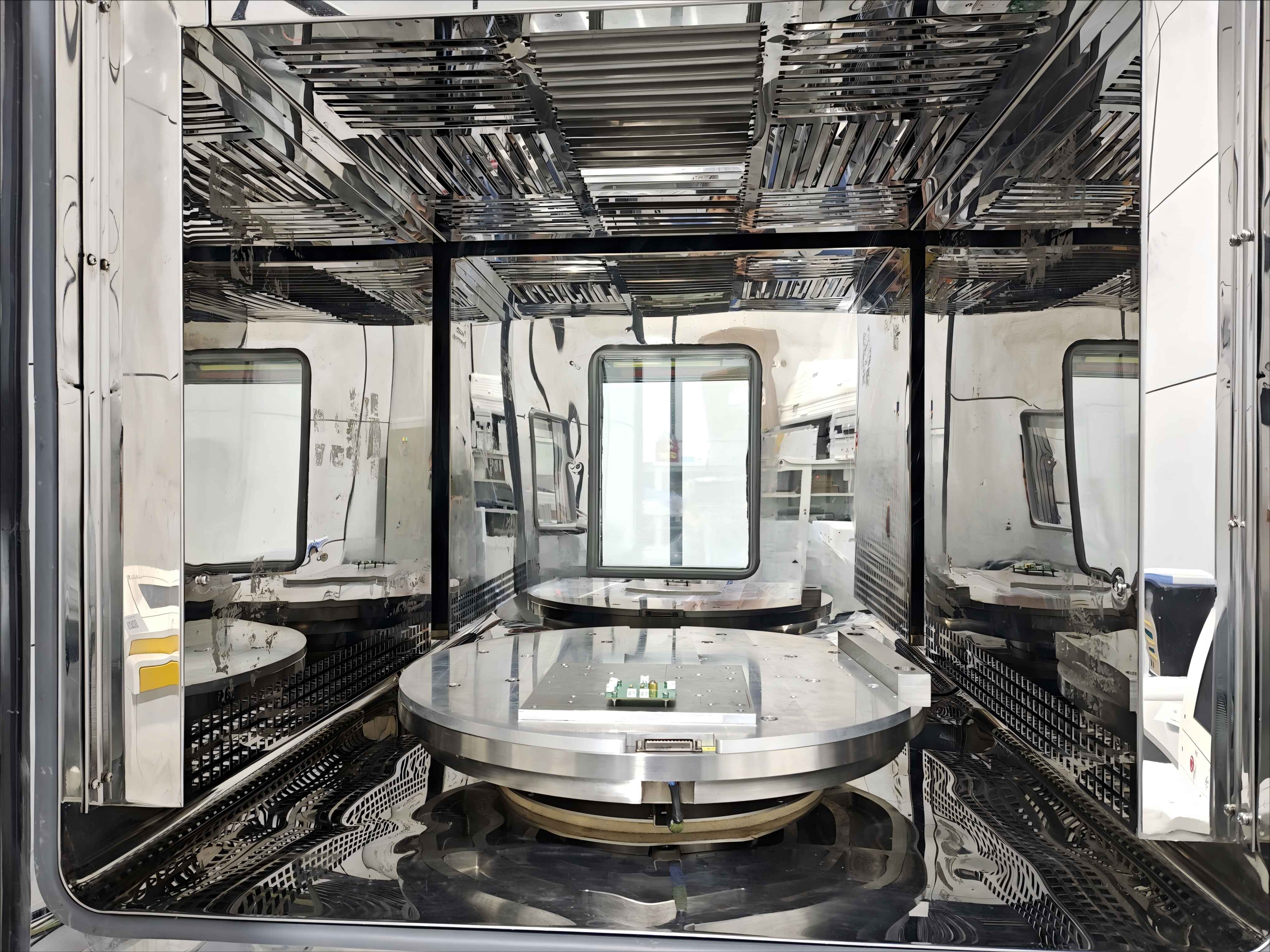}
            \caption{\label{thermal}}
        \end{subfigure}
	\caption{Space qualification test setup. (\subref{setting}) Micro-LED samples with interface plate, (\subref{vibration}) vibration table loaded with micro-LED samples, (\subref{shock}) shock table loaded with micro-LED samples, and (\subref{thermal}) thermal chamber for thermal cycling testing.}
	\label{space}
\end{figure}

With the photoelectric effects demonstrated for multiple wavelengths, the next question is whether the micro-LED is suitable for space applications, given its compact in size and weight. The space qualification process comprises several tests, among which the most relevant to the micro-LED are mechanical and thermal experiments that require each micro-LED to be characterized after every step. The mechanical experiments include sinusoidal vibration, random vibration and shock. The test setup is shown in Figure \ref{space}, including the micro-LED samples, vibration table, shock table and thermal chamber. The following relations of the micro-LEDs were gathered: V-I, I-P, and beam spectral profile. In this experiment, four wavelengths of micro-LEDs were tested, including 254nm, 262 nm, 274 nm and 282 nm. 

\subsection{Mechanical experiments}
\begin{figure}[H]
\centering
	\includegraphics[width=0.45\textwidth]{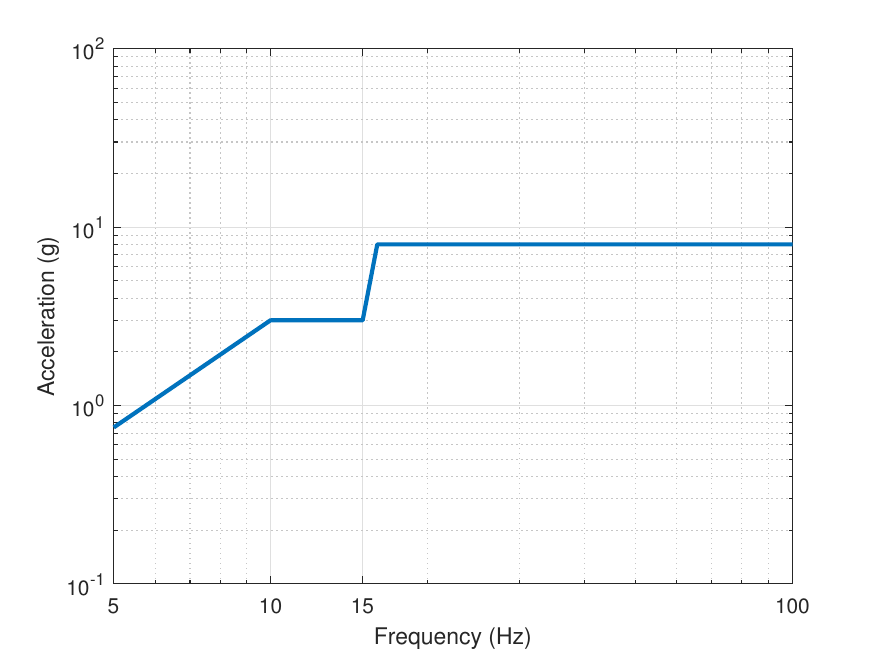}
	\caption{Sinusoidal vibration test conditions for micro-LED qualification testing.}
	\label{sinusoidalfig}
\end{figure}

\begin{figure}[H]
\centering
	\includegraphics[width=0.45\textwidth]{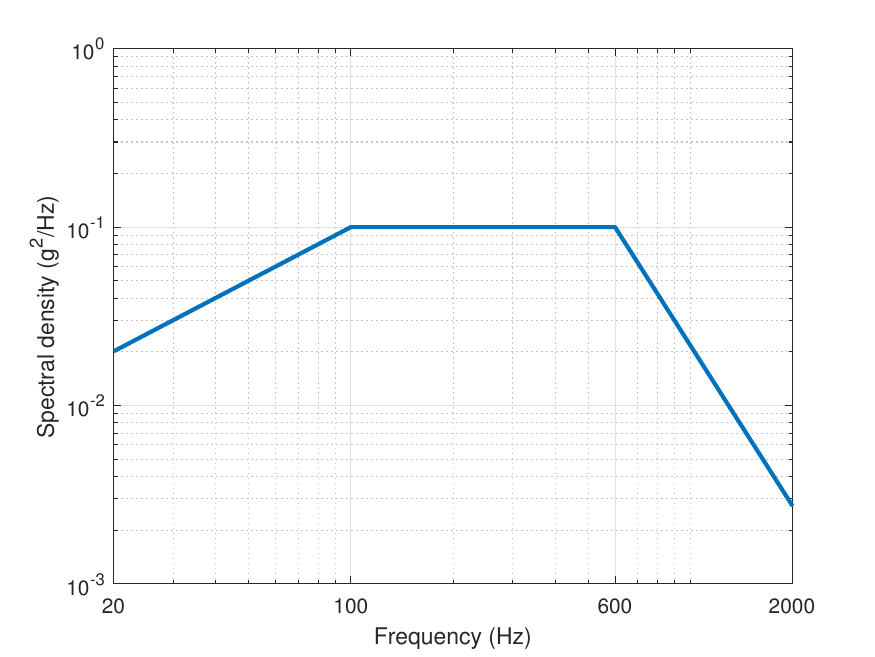}
	\caption{Random vibration spectral density of composite frequency spectrum.}
	\label{vibrationfig}
\end{figure}

The spacecraft is subject to a variety of complex mechanical environments during launch and in-orbit operation, For the mechanical tests of sinusoidal vibration and random vibration, the spectrum of sinusoidal vibration is shown is shown in Figure \ref{sinusoidalfig} and the parameters for the tests are shown in Table \ref{sinusoidaltab}. The spectrum of random vibration is shown in Figure \ref{vibrationfig} and reaches a total of 9.07 gRMS. The input parameters for the spectral envelope
are shown in Table \ref{vibrationtab}.

\begin{table}[H]
\centering
    \caption{Parameters for the sinusoidal vibration spectral envelope.}
    \begin{tabular}{cc}
    \hline
    Frequency(Hz) & Amplitude or acceleration\\
    \hline
    5-10 & 7.47mm\\
    10-15 & 3.0g\\
    16-100 & 8.0g\\
    \hline
    \end{tabular}
    \label{sinusoidaltab}
\end{table}

\begin{table}[H]
\centering
    \caption{Parameters for the random vibration spectral envelope.}
    \begin{tabular}{cc}
    \hline
    Frequency(Hz) & Spectral Density\\
    \hline
    20-100 & +3dB/oct\\
    100-600 & $0.1{\rm g^2/Hz}$\\
    600-2000 & -9dB/oct\\
    \hline
    \end{tabular}
    \label{vibrationtab}
\end{table}

\begin{figure}[H]
\centering
	\includegraphics[width=0.45\textwidth]{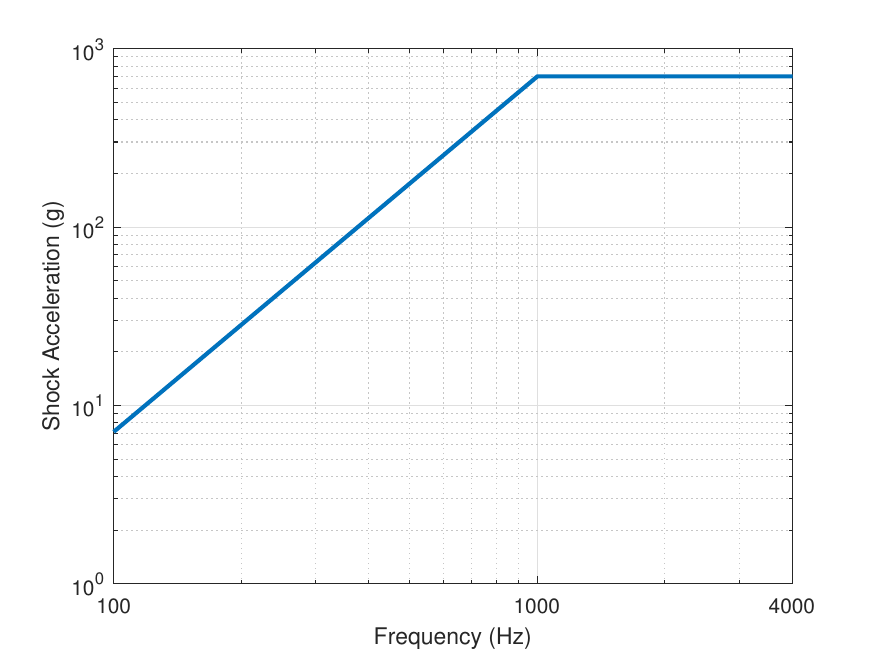}
	\caption{Shock profile spectrum used for micro-LED qualification testing.}
	\label{shockfig}
\end{figure}

During launch, significant shocks are exerted on the payload due to engine cutoff, stage separation, and payload release. To simulate this scenario on
ground, the micro-LED's shock test was performed, the shock profile is shown is shown in Figure \ref{shockfig} and the exact values are set as shown in Table \ref{shocktab}.

\begin{table}[H]
\centering
    \caption{Parameters for the spectrum of shock.}
    \begin{tabular}{cc}
        \hline
        Frequency(Hz) & Spectral Density\\
        \hline
        100-1000 & -9dB/oct\\
        1000-4000 & 700g\\
        \hline
    \end{tabular}
    \label{shocktab}
\end{table}

\subsection{Thermal experiments}

\begin{figure}[H]
\centering
	\includegraphics[width=0.45\textwidth]{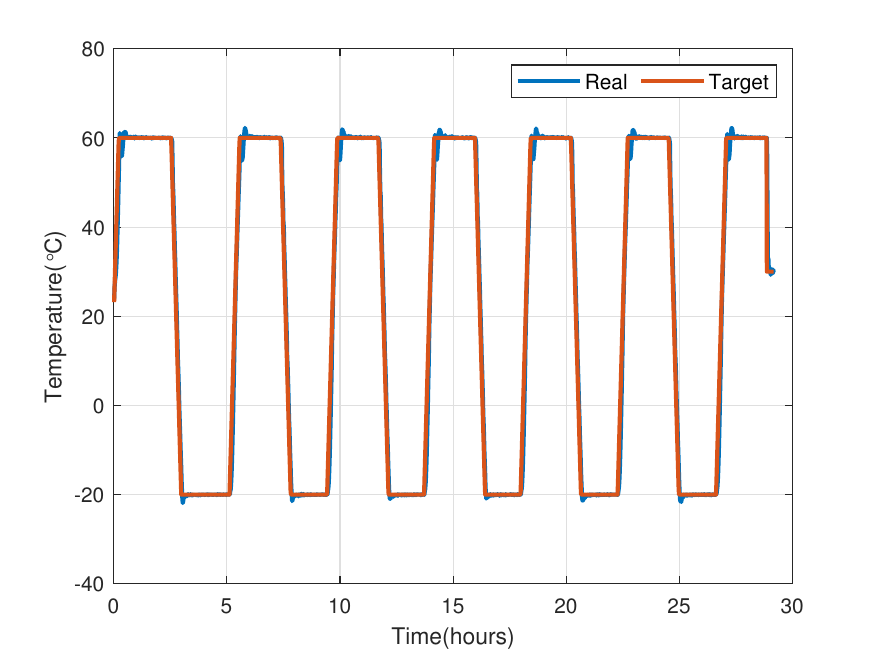}
	\caption{Time history of thermal cycling test in air.}
	\label{temperature}
\end{figure}

The micro-LEDs were subjected to the thermal cycling test, with a view
that micro-LED will operate in a near earth orbit with large temperature variation in future missions apart from the detection of gravitational waves in space. Preliminary a short-cycle test was run in air at ambient pressure.The target temperature and the measured  temperature in the thermal chamber is shown in Figure \ref{temperature}. The test ran for 6.5 cycles with temperature variation from -20 to +60 $^\circ$C.

The micro-LEDs were constantly switched on with 1 mA drive current and
run over a range of temperatures, as this will mimic the on orbit situations.

\subsection{Test Results}

\begin{figure}[H]
    \centering
        \begin{subfigure}{0.4\textwidth}
    	\includegraphics[width=0.9\textwidth]{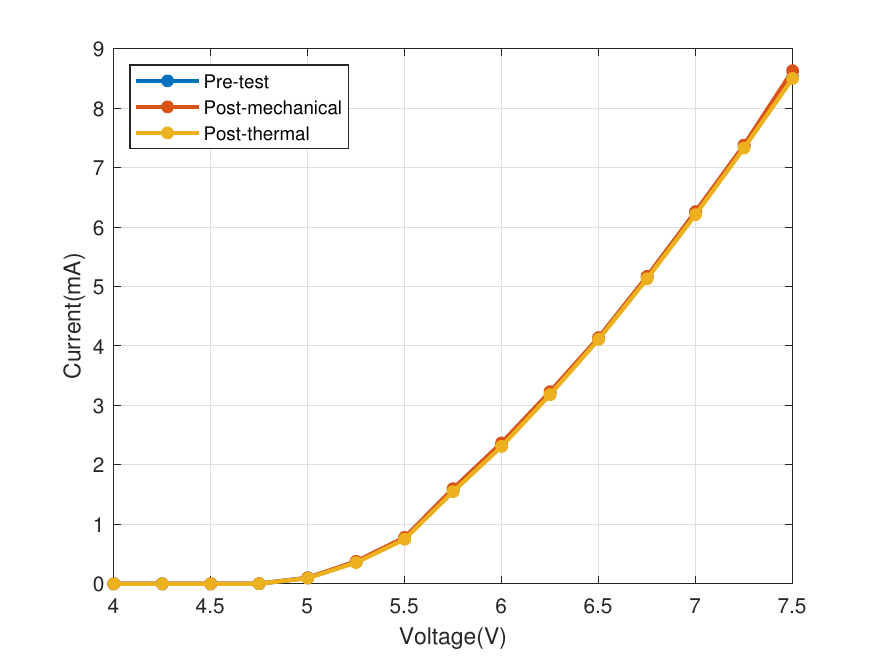}
            \caption{\label{255UI}}
        \end{subfigure}
        \begin{subfigure}{0.4\textwidth}
    	\includegraphics[width=0.9\textwidth]{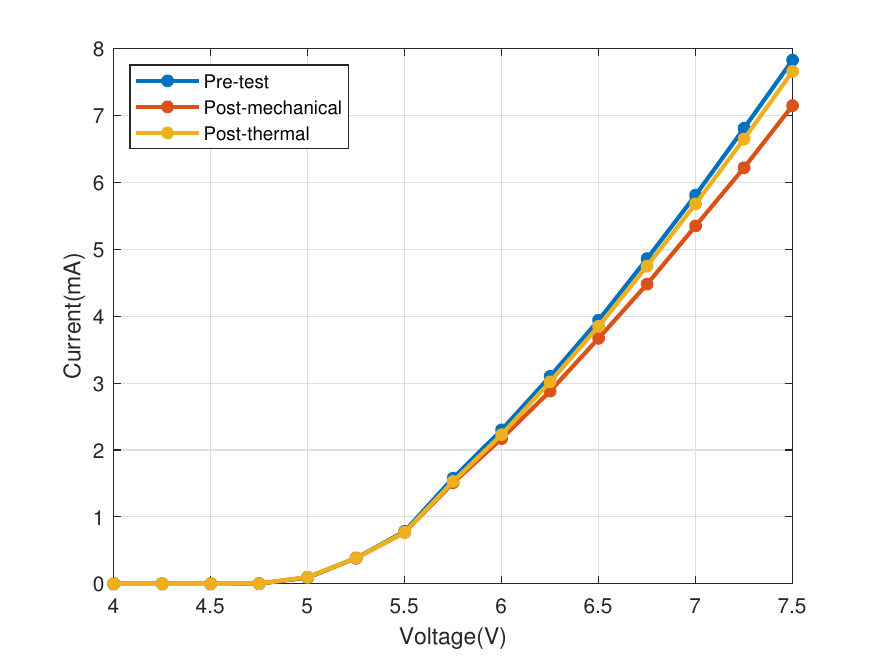}
            \caption{\label{265UI}}
        \end{subfigure}
        \begin{subfigure}{0.4\textwidth}
    	\includegraphics[width=0.9\textwidth]{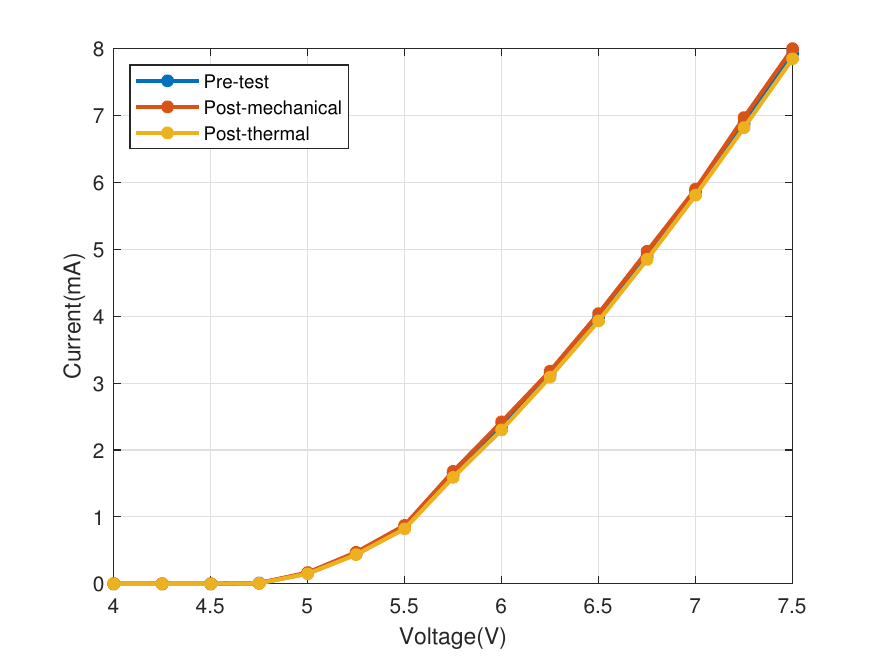}
            \caption{\label{275UI}}
        \end{subfigure}
        \begin{subfigure}{0.4\textwidth}
    	\includegraphics[width=0.9\textwidth]{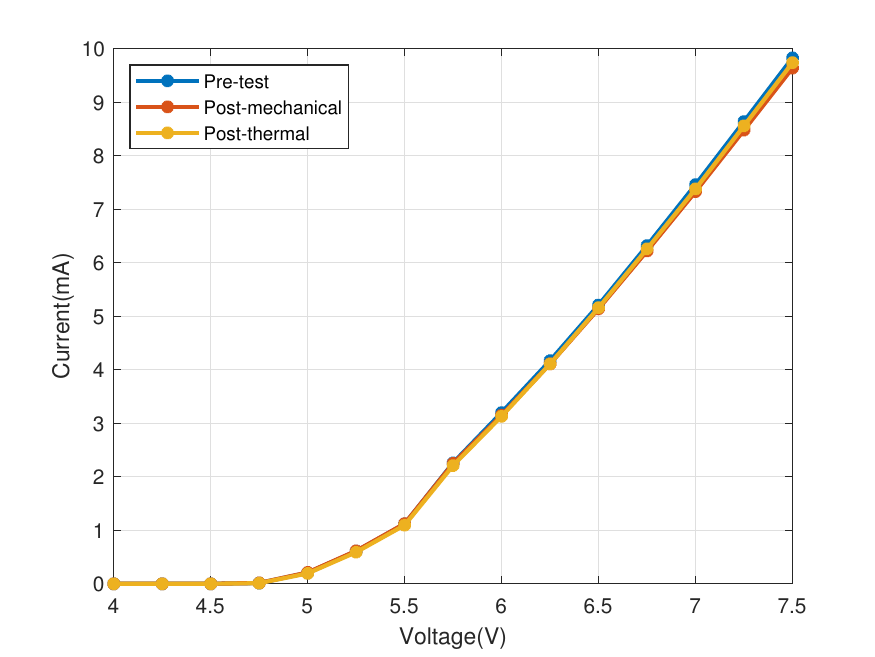}
            \caption{\label{285UI}}
        \end{subfigure}
	\caption{Voltage vs drive current of selected micro-LEDs before test, during, and after test. (\subref{255UI}) 254 nm, (\subref{265UI}) 262 nm, (\subref{275UI}) 274 nm, and (\subref{285UI}) 282 nm.}
	\label{UIMech}
\end{figure}

\begin{figure}[H]
    \centering
        \begin{subfigure}{0.4\textwidth}
    	\includegraphics[width=0.9\textwidth]{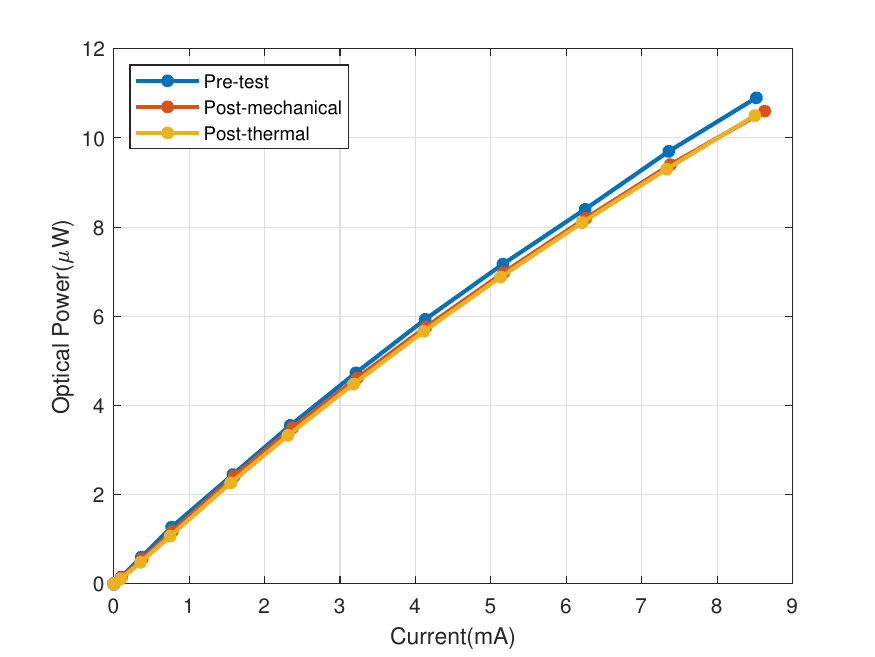}
            \caption{\label{255IP}}
        \end{subfigure}
        \begin{subfigure}{0.4\textwidth}
    	\includegraphics[width=0.9\textwidth]{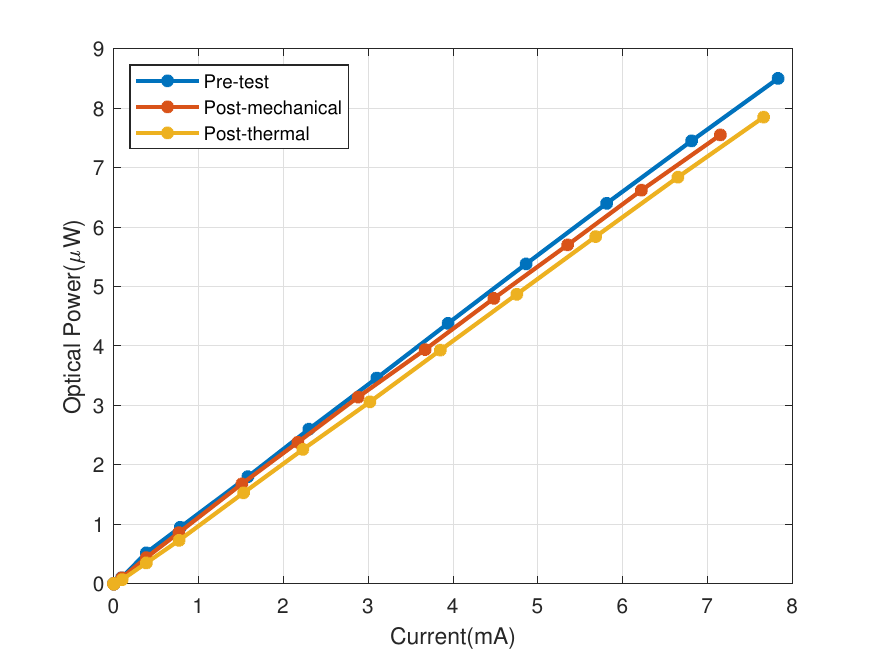}
            \caption{\label{265IP}}
        \end{subfigure}
        \begin{subfigure}{0.4\textwidth}
    	\includegraphics[width=0.9\textwidth]{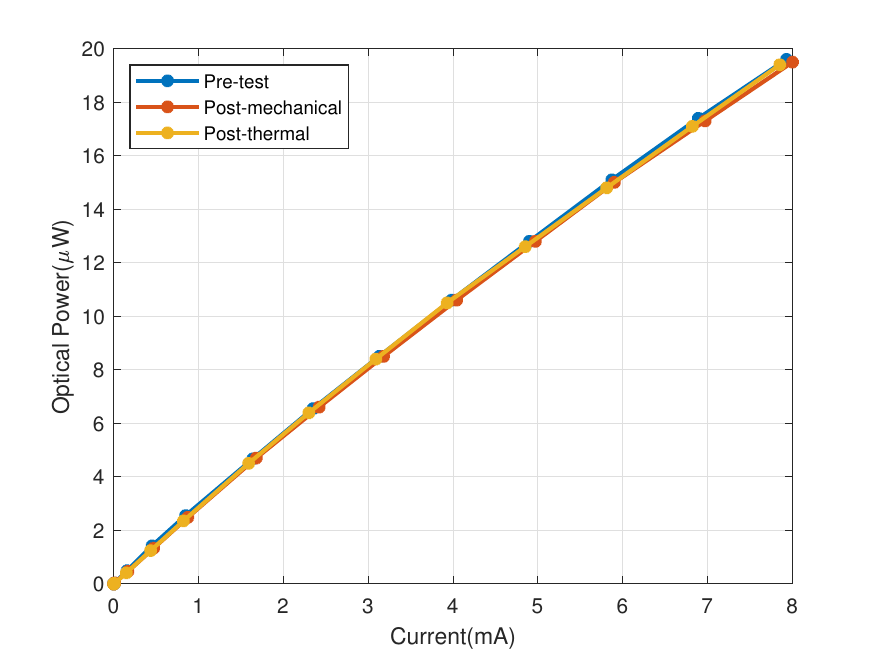}
            \caption{\label{275IP}}
        \end{subfigure}
        \begin{subfigure}{0.4\textwidth}
    	\includegraphics[width=0.9\textwidth]{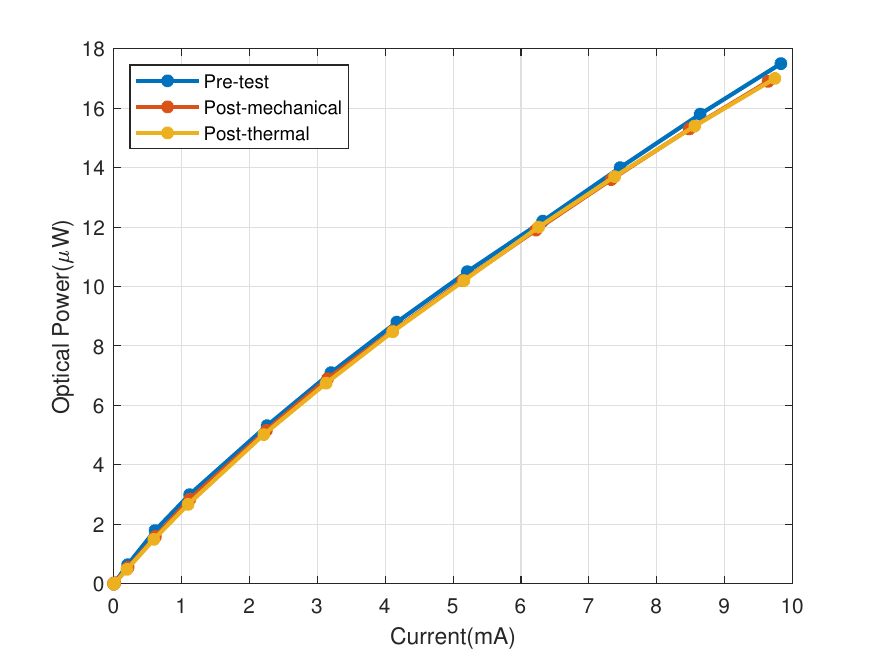}
            \caption{\label{285IP}}
        \end{subfigure}
	\caption{Optical power vs drive current of selected micro-LEDs before test, during, and after test. (\subref{255IP}) 254 nm, (\subref{265IP}) 262 nm, (\subref{275IP}) 274 nm, and (\subref{285IP}) 282 nm.}
	\label{UPMech}
\end{figure}

\begin{figure}[H]
    \centering
        \begin{subfigure}{0.4\textwidth}
    	\includegraphics[width=0.9\textwidth]{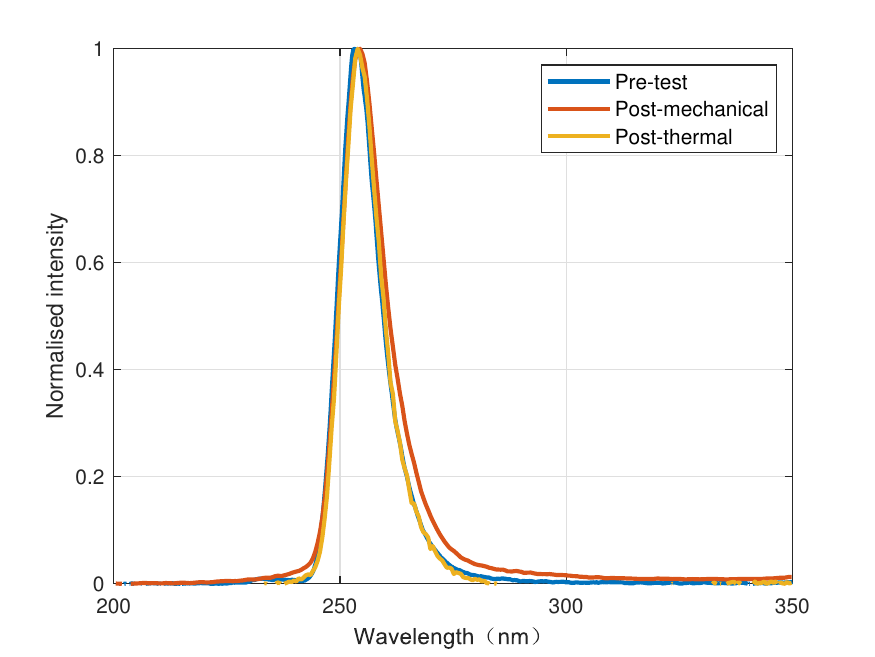}
            \caption{\label{255spectrum}}
        \end{subfigure}
        \begin{subfigure}{0.4\textwidth}
    	\includegraphics[width=0.9\textwidth]{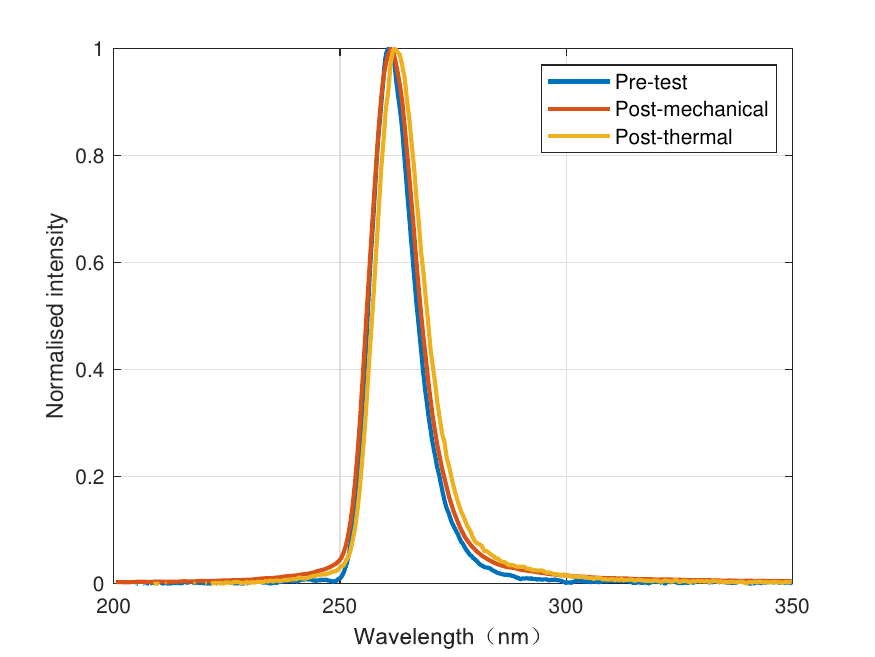}
            \caption{\label{265spectrum}}
        \end{subfigure}
        \begin{subfigure}{0.4\textwidth}
    	\includegraphics[width=0.9\textwidth]{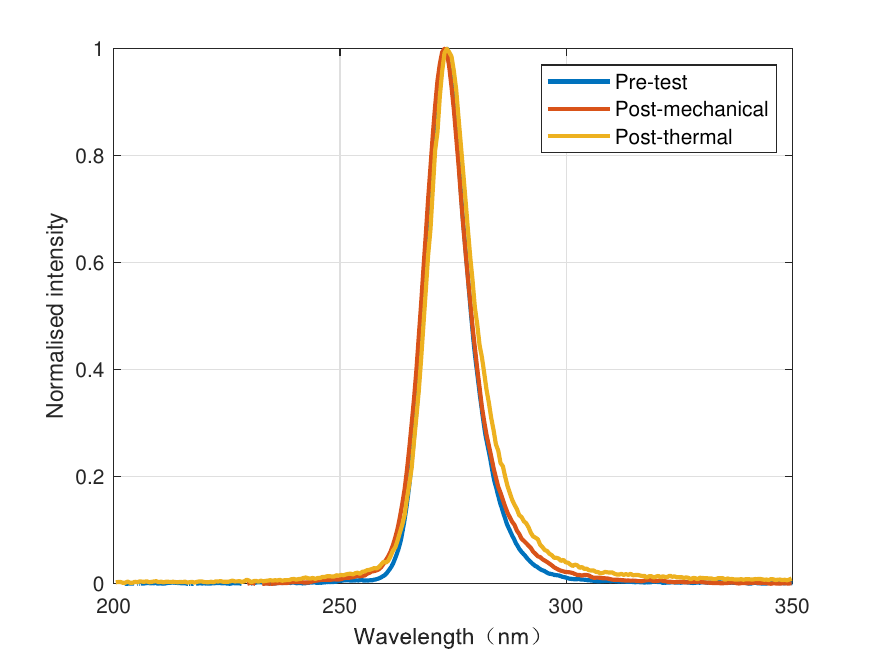}
            \caption{\label{275spectrum}}
        \end{subfigure}
        \begin{subfigure}{0.4\textwidth}
    	\includegraphics[width=0.9\textwidth]{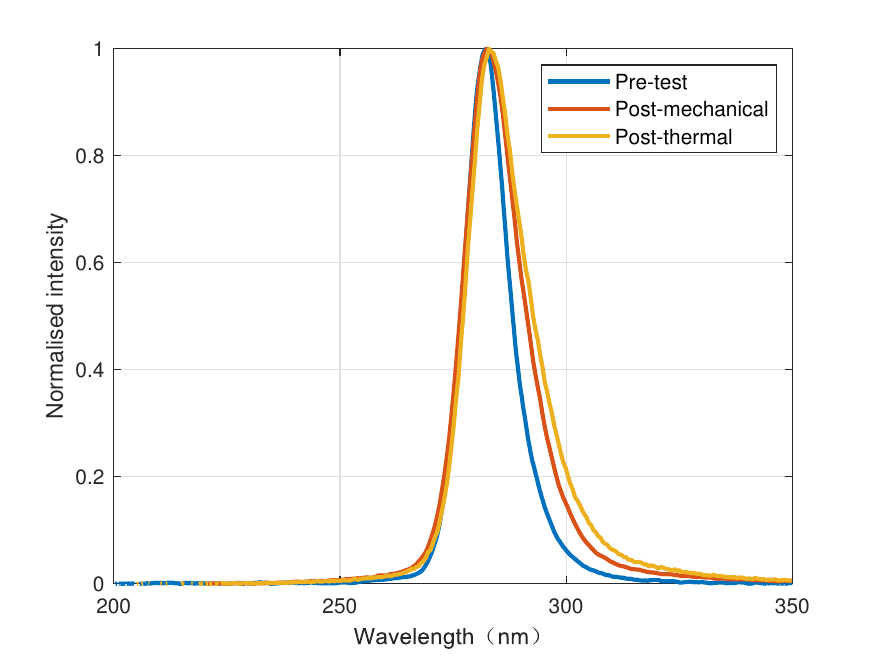}
            \caption{\label{285spectrum}}
        \end{subfigure}
	\caption{Emission spectrum of selected micro-LEDs before test, during, and after test. (\subref{255spectrum}) 254 nm, (\subref{265spectrum}) 262 nm, (\subref{275spectrum}) 274 nm, and (\subref{285spectrum}) 282 nm.}
	\label{SpectrumMech}
\end{figure}

The three key parameters of V-I curve, I-P curve, and spectrum were measured and recorded. A summary of parameters variation between pre-test and post-test from four micro-LEDs are shown in Table \ref{charaMech}. Detailed results are shown in Figure \ref{UIMech}-\ref{SpectrumMech}. 

\begin{table}[H]
\caption{Summary of parameters variation after space qualification tests.}
\centering
    \begin{tabular}{cccc}
    \hline
    Wavelength(nm)&$\Delta$ V-I&$\Delta$ I-P&$\Delta$ Peak wavelength\\
    \hline
    254&-1.5{\%}&-3.6{\%}&-0.20{\%}\\
    262&-1.2{\%}&-4.8{\%}&0.19{\%}\\
    274&-2.2{\%}&-0.33{\%}&-0.04{\%}\\
    282&-1.9{\%}&-2.4{\%}&0.18{\%}\\
    \hline
    \end{tabular}\\
    \label{charaMech}
\end{table}

The V-I curves variation of the tested micro-LEDs, shown in Figure \ref{UIMech}, ranged from -2.2{\%} to -1.2{\%}. The I-P curves in optical power all showed less than 5{\%} variation, shown in Figure \ref{UPMech}. The observed shift of the spectral peak was less than 1{\%}, shown in Figure \ref{SpectrumMech}.

These results indicate that the micro-LEDs have a fundamentally high reliability. The output optical power can be well controlled using current source from I-P curve. The peak wavelength remains unchanged and there is very little change in the output spectrum. This demonstrates that the photoelectric effect is not disturbed by the harsh environment.

The level of robustness indicates that micro-LEDs are well suited for
charge management system. Space qualification of micro-LEDs has increased the micro-LED device status to TRL-5 in the sense that the devices ``operated in a simulated operational environment''. Subject to further radiation and thermal tests, it will approach TRL-6.

\section{Conclusion.}

The present work will have served its purpose if we succeed in demonstrating that micro-LED is a viable alternative to UV LED in charge management in the detection of gravitational waves in space. Compared with UV LEDs, micro-LEDs have higher modulation bandwidth and can achieve better resolution in optical power. The compactness in size and weight enable us to contemplate the prospect of integrating the micro-LED into the structure of an inertial sensor, without the need to use an optical fiber to direct light into the electrode housing.

Apart from the standard 255 nm wavelength, micro-LEDs with multiple
wavelengths were studied experimentally. This will provide useful insights in the system design of the open loop control charge management system, especially when solar activities during different phases of a solar cycle are considered.

A key part of our work is to space qualify the micro-LED and raise its
technological readiness. With further radiation and thermal tests and integration into an inertial sensor, the micro-LED is ready to be further tested in space and thereby provide an alternative charge management system for future gravitational wave missions. It is also conceivable that the more compact charge management system may be useful for future experimental test of general relativity in space when a lighter payload for inertial sensor is called for. More work remains to be done to further explore this option in our future work. 

\section*{Acknowledgments}

\subsection*{Funding}
This work was supported by the National Key R{\&}D Program of China (Task No. 2021YFC2202500).

\subsection*{Author Contributions} 
Yun Kau Lau, Zongfeng Li, Hongqing Huo and Pengfei Tian participated in the design and interpretation of the reported experiments or results. Yuandong Jia, Zhihao Zhang, Yinbowen Zhang, Yuning Gu and Zongfeng Li participated in the acquisition and/or analysis of data. Yuandong Jia, Yinbowen Zhang, Suwen Wang, Pengfei Tian, Zongfeng Li and Yun Kau Lau participated in drafting and/or revising the manuscript. Yuandong Jia, Yinbowen Zhang, Zemin Zhang and Yi Zhang contributed to the experimental data analysis. Guozhi Chai, Shanduan Zhang, Pengfei Tian, Hongqing Huo and Zongfeng Li provided administrative, technical or supervisory support. 

\subsection*{Conflicts of Interest}
The authors declare that there is no conflict of interest regarding the publication of this article.

\subsection*{Data Availability}
The data used to support the findings of this study are available from the corresponding author upon reasonable request.

\printbibliography

\end{document}